\documentclass[seceq,preprint]{ptptex}
\usepackage{amsmath,amssymb,mathtools}
\usepackage{graphicx}

\newcommand{\mspd}[1]{[#1]_G}
\newcommand{\epsi}{(\eta\Psi)}
\newcommand{\qpsi}{(Q_G\Psi)}
\newcommand{\qxi}{(Q_G\Xi)}
\newcommand{\eLambda}{\eta\Lambda_1}
\newcommand{\qLambda}{Q_G\Lambda_{\frac{1}{2}}}
\newcommand{\eLambdaf}{\eta\Lambda_{\frac{3}{2}}}

\newcommand{\smallcomb}[2]{(\begin{smallmatrix}
		    #1\\#2
		   \end{smallmatrix})}
\newcommand{\comb}[2]{\left(\begin{array}{c}
		    #1\\#2
		   \end{array}\right)}

\notypesetlogo                       
\preprintnumber[3cm]{
YITP-13-131\\ 
}

\title{
The Ramond Sector of Heterotic String Field Theory
}

\author{
Hiroshi \textsc{Kunitomo}\footnote{%
E-mail:\  {\tt kunitomo@yukawa.kyoto-u.ac.jp}}
}

\inst{
Yukawa Institute for Theoretical Physics, Kyoto University, \\
Kyoto 606-8502, Japan
}

\abst{
We attempt to construct the full equations of motion 
for the Neveu-Schwarz and the Ramond sectors of 
the heterotic string field theory.
Although they are also non-polynomial in the Ramond string field $\Psi$, 
we can construct them order by order in $\Psi$.
Their explicit forms with the gauge transformations are given
up to the next-to-next-to-leading order in $\Psi$.
We also determine a subset of the terms to all orders. 
By introducing an auxiliary Ramond string field $\Xi$, 
we construct a covariant action supplemented with a constraint,
which should be imposed on the equations of motion.
We propose the Feynman rules and show how they reproduce
well-known physical four-point amplitudes with external fermions.
}

\begin{document}

\maketitle

\section{Introduction}

The Neveu-Schwarz (NS) sector of the heterotic string field theory 
was first constructed order by order in the coupling constant 
using the large Hilbert space of the superghosts.\cite{Okawa:2004ii} 
Soon afterwards, it was completed by giving the action and gauge transformations in a closed form 
as a Wess-Zumino-Witten (WZW)-like theory,\cite{Berkovits:2004xh} 
which is a natural extension of the similar formulation of the open superstring 
field theory\cite{Berkovits:1995ab}. This heterotic string field theory is 
a closed string field theory with non-polynomial 
interactions (in the NS string field), which can be constructed based on the polyhedral 
overlapping conditions.\cite{Nonpolynomial,Nonpolynomial2,Nonpolynomial3,KKS,Kugo:1989tk,Zwiebach:1992ie}
The corresponding string products satisfy a set of identities 
with an additional derivation $\eta$ characterizing the algebraic structure
of the heterotic string field theory. An analog of the Maurer-Cartan form, 
the basic ingredient of the WZW-like action, can be constructed 
by means of the pure gauge field in bosonic closed string field theory.\cite{Berkovits:2004xh}

On the other hand, the Ramond (R) sector of the heterotic string 
field theory has not been studied so far. This is presumably 
because one of the main motivations to construct string field theories 
is to find classical solutions, including the solution for tachyon condensation, 
which does not require the R sector. 
Nevertheless, it is important to understand the R sector for many other
purposes, such as discussing non-renormalization theorems, supersymmetry breaking,
and so on. The aim of this paper is to attempt to construct the full heterotic
string field theory by including the R sector.

In the case of the WZW-like open superstring field theory,
however, it was not straightforward to construct the covariant 
R action due to the picture number mismatch.
Therefore, at first, only the equations of motion could be given
in the covariant form.\cite{Berkovits:2001im} 
Subsequently, a covariant action was constructed by introducing
an auxiliary R string field with a constraint.\cite{Michishita:2004by}
The auxiliary string field is eliminated by the constraint,
and the equations of motion reduce to those previously obtained.
According to this open superstring case,
we will at first attempt to construct the equations of motion 
for the NS and the R sectors of the heterotic string field theory 
since the picture number difficulty is common 
to the heterotic string case. 

From simple consideration, one can easily notice that 
the full equations of motion in the heterotic string field theory
have to be non-polynomial not only in the NS string field but 
also in the R string field to reproduce correct fermion amplitudes. 
We will construct them order by order in the R string field $\Psi$
by imposing some consistency conditions, including invariance 
under a gauge transformation. Their explicit forms will be given
up to the next-to-next-to-leading order corrections in $\Psi$. 
The non-polynomial gauge transformations will also be determined 
up to the same order. In addition, we will find a subset of terms 
in the equations of motion and the gauge transformations built 
with one or two string products to all orders.

Then we will introduce an auxiliary R string field and 
construct an action.
The above equations of motion can be obtained from those derived from 
the action by eliminating the auxiliary field using 
a constraint,\cite{Michishita:2004by} which is also non-polynomial in $\Psi$. 
We will also propose the Feynman rules to compute tree amplitudes.
Since the action has to be supplemented by the constraint, 
these Feynman rules cannot be simply derived 
by the conventional method. Nevertheless, we can extend the rules 
proposed for the open superstring\cite{Michishita:2004by} 
to those of the heterotic string with
a few additional prescriptions. As evidence that these Feynman rules 
work well, we will show that they actually reproduce
on-shell four-point amplitudes with external fermions.

This paper is organized as follows.
In \S \ref{WZW}, we will briefly summarize the known results on 
the NS sector of the heterotic string field theory. 
The heterotic string fields are defined based on the large Hilbert 
space of superghosts for the holomorphic, superstring, sector. 
Together with the NS string field $V$, we will also introduce 
the R string field $\Psi$ here.
After explaining the fundamental properties of the string products,
the WZW-like action of the NS sector will be given.
We will attempt to construct the full equations of motion, 
including interactions with the R sector, in \S\ref{eom including R sector},
and find a general form of the full equations of motion required 
by a gauge invariance.
The explicit forms of the equations of motion and the gauge transformations 
will be given up to the next-to-next-to-leading order in $\Psi$. 
We will also determine a subset of terms to all orders. 
We will construct, in \S\ref{action with a constraint},
an action and a constraint by introducing an auxiliary R string field $\Xi$.
In \S\ref{Feynman rules}, we will propose the Feynman rules
by adding a few prescriptions to the straightforward extensions of 
those for the open superstring field theory.\cite{Michishita:2004by}
We will show how well-known four-point fermion amplitudes
are reproduced. 
Section \ref{Discussion} is devoted to open questions and discussion. 
We add three appendices.  In Appendix \ref{appendix A}, 
We will give the explicit forms of the gauge transformations 
at the next-to-next-to-leading order in $\Psi$. 
We will explain in detail how the general form of 
the full equations of motion are obtained by imposing a gauge invariance
in Appendix \ref{appendix B}.
The derivation of the all-order 
results in \S\ref{eom including R sector} will be presented 
in Appendix \ref{appendix C}. The gauge transformation laws 
built with two string products will also be given.

\section{The NS sector of the heterotic string field theory}\label{WZW}

In this section, we summarize the known results in the NS sector of
the heterotic string field theory.\cite{Okawa:2004ii,Berkovits:2004xh}. 
Together with the NS string field $V$, the R string field $\Psi$ is also
introduced. After introducing the string products 
and discussing their fundamental properties, the gauge-invariant 
WZW-like action is given.

\subsection{Heterotic string fields}

The first quantized heterotic string  in a consistent background
is in general described by a conformal field theory with the following
structure of holomorphic and anti-holomorphic sectors.
The holomorphic sector is composed of the Hilbert spaces of an $N=1$ 
superconformal matter with central charge $c=15$, the reparametrization 
ghosts $(b(z),c(z))$ with $c=-26$ and the bosonized superghosts 
$(\xi(z), \eta(z), \phi(z))$ with $c=11$. In particular, we adopt 
the large Hilbert space for the superconformal ghosts involving
the zero mode $\xi_0$ of the field $\xi(z)$.
The anti-holomorphic sector consists of a conformal matter theory 
with $\bar{c}=26$ and the reparametrization ghosts 
$(\bar{b}(\bar{z}),\bar{c}(\bar{z}))$ with $\bar{c}=-26$. 
The normalization of correlation function in this Hilbert space is given by
\begin{equation}
 \langle\langle \xi e^{-\phi} c\partial c\partial^2 c
\bar{c}\bar{\partial}\bar{c}\bar{\partial}^2\bar{c}
\rangle\rangle =2.
\end{equation}
There are two sectors, the NS sector and the R sector, 
in the full heterotic string theory depending on the boundary conditions 
of the holomorphic world-sheet fermions. The states in the NS (R) sector 
represent space-time bosons (fermions). 

Using the BRST operator $Q$, the linearized equations of motion are given by
\begin{subequations}\label{linearized eom}
 \begin{align}
 Q\eta V =& 0,\label{linearized eom NS}\\
 Q\eta\Psi =& 0,\label{linearized eom R}
\end{align}
\end{subequations}
for the NS string field $V$ and the R string field $\Psi$, respectively.
In this paper, we denote the zero-mode $\eta_0$ of the field $\eta(z)$ 
as $\eta$, for simplicity.
Due to the nilpotency $Q^2=\eta^2=0$ and the anti-commutativity $\{Q,\eta\}=0$,
these equations of motion are invariant under the gauge transformations
\begin{subequations}\label{linearized gauge tf} 
\begin{align}
 \delta V =& Q\Lambda_0+\eta \Lambda_1,\label{linearized gauge tf 1}\\
 \delta \Psi =&Q\Lambda_{\frac{1}{2}}+\eta \Lambda_{\frac{3}{2}},
\label{linearized gauge tf 2}
\end{align}
\end{subequations}
where $\Lambda_n$ with integer (half-integer) $n$ are the gauge 
parameter NS (R) string fields with the picture number $n$ as shown below. 
The on-shell physical states obtained from
these equations of motion and gauge transformations coincide with 
those given by the conventional BRST cohomology in the small
Hilbert space.\cite{Berkovits:1994vy} 

In the NS string field $V$, a  physical vertex operator takes the form of 
\begin{equation}
|V\rangle=\cdots+\xi_0 e^{-\phi}c\bar{c}V_A^{(NS)}(k)|0\rangle\varphi^A(k)+\cdots,
\label{NS string field}
\end{equation}
where $V_A^{(NS)}$ is a general matter primary operator with conformal dimensions $(1/2,1)$.
The coefficient $\varphi^A(k)$ represents a space-time boson field, where 
the space-time tensor indices are simply denoted as $A$.
Since the matter vertex  $V_A^{(NS)}$ is Grassmann odd,\footnote{
In this paper, we focus on the supersymmetric theory and assume that 
the NS states are projected onto the GSO even sector.}
the NS string field $V$ is found to be Grassmann odd.
We also find, from the structure of (\ref{NS string field}), that 
the NS string field $V$ has the ghost number $(G)$ equal to one and 
the picture number $(P)$ equal to zero. In summary,
\begin{equation}
 V\ :\ \textrm{Grassmann\ odd},\ (G,P)=(1,0).
\end{equation}
Similarly, a physical vertex operator in the  R string field 
has the form
\begin{equation}
|\Psi\rangle=\cdots +\xi_0 e^{-\phi/2}c\bar{c}V_\alpha^{(R)}(k)|0\rangle\psi^\alpha(k)+\cdots. 
\label{R string field}
\end{equation}
The matter primary operator $V_\alpha^{(R)}$ has conformal dimension $(5/8,1)$,
whose holomorphic sector is including the matter spin operator and $\alpha$ 
denotes the space-time tensor-spinor indices. 
We can assign odd Grassmann parity to the vertex $e^{-\phi/2}V_\alpha^{(R)}$
after imposing the GSO projection. 
Taking account of the fact that the space-time fermion fields $\psi^\alpha(k)$ are 
Grassmann odd, we can read from (\ref{R string field}) that 
the R string field $\Psi$ is Grassmann odd and has the ghost number one
and the picture number one-half:
\begin{equation}
 \Psi\ :\ \textrm{Grassmann\ odd},\ (G,P)= (1,\frac{1}{2}).
\end{equation}
The gauge parameter string fields $\Lambda_n$ are found, from (\ref{linearized gauge tf}),
to be Grassmann even and have $(G,P)=(0,n)$.

All the heterotic string fields and the parameters,$V$, $\Psi$, and $\Lambda_n$ 
also satisfy the subsidiary conditions for the closed string:
\begin{equation}
b_0^-V  = L_0^-V = 0,\qquad 
b_0^-\Psi = L_0^-\Psi = 0,\qquad
b_0^-\Lambda_n = L_0^-\Lambda_n = 0,
\end{equation}
where
\begin{equation}
c_0^\pm = \frac{1}{2}(c_0\pm\bar{c}_0),\qquad
b_0^\pm = b_0\pm \bar{b}_0,\qquad L_0^\pm = L_0\pm\bar{L}_0. 
\end{equation}
We also define an inner product by $\langle A,B\rangle = \langle A|c_0^-|B\rangle$,
where $\langle A|$ is the BPZ conjugate state of $|A\rangle$. This inner product satisfies
\begin{align}
\langle A,B\rangle =& (-1)^{(A+1)(B+1)}\langle B,A\rangle, \\
\langle XA,B\rangle =& (-1)^A\langle A,XB\rangle,\ \textrm{with}\
X= Q\ \textrm{or}\ \eta,
\end{align}
and is non-zero if and only if the total ghost and picture numbers are equal to
four and minus one, respectively:
\begin{equation}
 G(A)+G(B) = 4,\qquad P(A)+P(B)=-1.
\end{equation}  

\subsection{String products and identities}\label{algebra}

As in the bosonic string field theory,\cite{Kugo:1989tk, Zwiebach:1992ie} 
the $n$ string product $(n\ge2)$ in the heterotic string field theory
is defined by a multilinear map from $n$ string fields $\{B_i\}$ 
$(i=1,\cdots,n)$ 
annihilated by $b_0^-$ and $L_0^-$ to one string field $[B_1,\cdots,B_n]$
also annihilated by $b_0^-$ and $L_0^-$. 
Each string field $B_i$ can either be an NS string field or an R string field. 
We can find that the resultant string field $[B_1,\cdots,B_n]$ is an NS (R) string field
if the number of the R string in $\{B_i\}$ is even (odd) from the cut structure of 
the world-sheet fermions.\cite{Hata:1987qx}
This is also consistent with the space-time fermion number conservation.
The string field $[B_1,\cdots,B_n]$ has ghost and picture numbers 
\begin{equation}
(G,P)=\left(\sum_{i=1}^n G_i-2n+3,\ \sum_{i=1}^n P_i\right), \label{g and p}
\end{equation}
where $(G_i,P_i)$ are those of $B_i$.

The BRST operator $Q$ does not act as derivation on these string products but 
satisfies\cite{Zwiebach:1992ie} 
\begin{align}
 0 =& Q[B_1,\cdots,B_n]+\sum^n_{i=1}(-1)^{(B_1+\cdots+B_{i-1})}[B_1,\cdots,QB_i,\cdots,B_n]\nonumber\\
&+\sum_{\underset{l+k=n}{\{i_l,j_k\}}}\sigma(i_l,j_k)[B_{i_1},\cdots,B_{i_l},[B_{j_1},\cdots,B_{j_k}]],
\label{main}
\end{align}
where $\sigma(i_l,j_k)$ is a sign factor defined to be the sign picked up when one
rearranges the sequence $\{Q,B_1,\cdots,B_n\}$ into the order 
$\{B_{i_1},\cdots,B_{i_l},Q,B_{j_1},\cdots,B_{j_k}\}$.
These identities have exactly the same form as those of bosonic closed 
string field theory because they have the same structure of the anti-ghost 
$b$ insertion. We have three additional derivations 
$X=\{\eta,\delta,\partial_t\}$ playing important roles
in the heterotic string field theory. They are graded commutative 
with the BRST operator $XQ=(-1)^XQX$ and act on the string products as derivations:
\begin{equation}
X[B_1,\cdots,B_n] = \sum^n_{i=1}(-1)^{X(1+B_1+\cdots+B_{i-1})}[B_1,\cdots,X B_i,\cdots,B_n].
\end{equation}

In addition to these fundamental BRST operator $Q$ and string products $[\cdots]$, 
it is useful to introduce a new BRST operator $Q_{B_0}$,
\begin{equation}
 Q_{B_0}B \equiv\ QB+\sum^\infty_{m=1}\frac{\kappa^m}{m!}[B_0^m,B],\label{shiftedQ}
\end{equation}
and string products $[\cdots]_{B_0}$,
\begin{equation}
[B_1,\cdots,B_n]_{B_0} \equiv \ \sum^\infty_{m=0}\frac{\kappa^m}{m!}[B_0^m,B_1,\cdots,B_n],
\label{shifted string products}
\end{equation}
shifted by a Grassmann even NS string field $B_0$ with $(G,P)=(2,0)$.
If $B_0$ satisfies the equation
\begin{equation}
\mathcal{F}(B_0) \equiv  QB_0+\sum_{n=2}^\infty\frac{\kappa^{n-1}}{n!}[B_0^n] = 0,\label{BEOM}
\end{equation}
the shifted BRST operator $Q_{B_0}$ is nilpotent and the shifted string products $[\cdots]_{B_0}$ 
satisfy the same form of identities as (\ref{main}):
\begin{align}
 0 =& Q_{B_0}[B_1,\cdots,B_n]_{B_0}
+\sum^n_{i=1}(-1)^{(B_1+\cdots+B_{i-1})}[B_1,\cdots,Q_{B_0}B_i,\cdots,B_n]_{B_0}\nonumber\\
&+\sum_{\underset{l+k=n}{\{i_l,j_k\}}}\sigma(i_l,j_k)[B_{i_1},\cdots,B_{i_l},[B_{j_1},\cdots,B_{j_k}]_{B_0}]_{B_0}.
\end{align}
The operators $X=\{\eta,\delta,\partial_t\}$ are, in contrast, neither graded
commutative with $Q_{B_0}$ nor derivation on the shifted string products $[\cdots]_{B_0}$.
Extra terms appear in the relations from the contribution that $X$ operates on $B_0$:
\begin{align}
& Q_{B_0}(XB)-(-1)^XX(Q_{B_0}B) = -\kappa[XB_0,B]_{B_0},\\
& X[B_1,\cdots,B_n]_{B_0}
=\sum_{i=1}^n(-1)^{X(1+B_1+\cdots+B_{i-1})}[B_1,\cdots,XB_i,\cdots,B_n]_{B_0}\nonumber\\
&\hspace{30mm}
+(-1)^X\kappa[XB_0,B_1,\cdots,B_n]_{B_0}.
\end{align}

\subsection{WZW-like action for the NS sector of the heterotic string}

As in the conventional WZW theory,
we introduce an extra dimension parameterized by $t\in [0,1]$.
The string field $V$ is extended 
to the $t$-dependent one $\mathcal{V}(t)$ satisfying $\mathcal{V}(0)=0$ 
and $\mathcal{V}(1)=V$. The fundamental quantities to construct the action
are the four string fields $B_Q(\mathcal{V})$ and $B_X(\mathcal{V})$ 
with $X=\{\eta,\delta,\partial_t\}$, which are analogs of 
the Maurer-Cartan form $g^{-1}dg$.
The most important one $B_Q(\mathcal{V})$ is defined by
the pure gauge field $G(\mathcal{V})$ in closed bosonic string field theory 
associated with a finite gauge parameter $\mathcal{V}$, which satisfies 
Eq.~(\ref{BEOM}): $\mathcal{F}(G)=0$.\cite{Berkovits:2004xh}
The remaining three, $B_X$, are determined so that they satisfy 
an analog of the Maurer-Cartan equation,
\begin{equation}
 XB_Q = (-1)^XQ_{B_Q}B_X.\label{MCequation}
\end{equation}
Although we can obtain the explicit forms of these basic quantities 
by solving some differential equations
order by order in $\kappa$,\cite{Berkovits:2004xh}
here we give only their first few terms as
\begin{align}
B_Q(\mathcal{V}) =& G(\mathcal{V})\nonumber\\
=& Q\mathcal{V}
+\frac{\kappa}{2}[\mathcal{V},Q\mathcal{V}]
+\frac{\kappa^2}{3!}([\mathcal{V},(Q\mathcal{V})^2]+[\mathcal{V},[\mathcal{V},Q\mathcal{V}]])+\cdots,\nonumber\\
B_X(\mathcal{V}) =& X\mathcal{V} 
+ \frac{\kappa}{2}[\mathcal{V},X\mathcal{V}]
+\frac{\kappa^2}{3!}(2[\mathcal{V},Q\mathcal{V},X\mathcal{V}]+[\mathcal{V},[\mathcal{V},X\mathcal{V}]])+\cdots,
\label{explicit Bs}
\end{align} 
which are enough for later use.

Using these \lq\lq Maurer-Cartan forms,\rq\rq 
the NS sector action of the heterotic string field theory
is written in the form\cite{Berkovits:2004xh}
\begin{equation}
 S_{NS} = \int^1_0 dt\langle\eta B_{\partial_t}, B_Q\rangle.\label{NS action}
\end{equation}
From Eq.~(\ref{MCequation}),
we can show that an arbitrary variation of this action can be written as
\begin{equation}
 \delta S_{NS} = -\langle \bar{B}_\delta,\eta\bar{B}_Q\rangle,\label{varS}
\end{equation}
where the string fields with a bar denote those evaluated at $t=1$, such as
$\bar{B}_\delta=B_\delta(\mathcal{V}(1))=B_\delta(V)$. In particular, we denote hereafter
$\bar{B}_Q=G(V)\equiv G$, for simplicity. As a result, the equation of motion becomes
\begin{equation}
 \eta G = 0.\label{NSEOM}
\end{equation}
By use of (\ref{varS}), one can also show that
the action (\ref{NS action}) is invariant under the gauge transformations,
\begin{equation}
 \bar{B}_\delta(V) = Q_G\Lambda_0 + \eta\Lambda_1,\label{NS gauge tf}
\end{equation}
which are the non-linear extensions of (\ref{linearized gauge tf 1}).\footnote{
Hereafter,  the gauge transformations of $V$ are given in the form of $\bar{B}_\delta(V)$,
which can be solved for $\delta V$ order by order in $\kappa$.\cite{Berkovits:2004xh}}
Here the operator $Q_G$ is the nilpotent BRST operator (\ref{shiftedQ}) shifted by the $G(V)$.
One can easily find that the equation of motion (\ref{NSEOM}) and gauge transformations 
(\ref{NS gauge tf}) coincide with (\ref{linearized eom NS}) and 
(\ref{linearized gauge tf 1}), respectively, in the lowest order in $\kappa$.

We note that the string field $\eta G(V)$ in the left-hand side of the equation of
motion (\ref{NSEOM}) satisfies two identities,
\begin{align}
 Q_G(\eta G)\equiv 0,\label{id1}\\
 \eta(\eta G)\equiv 0,\label{id2}
\end{align}
following from (\ref{MCequation}) and the nilpotency $\eta^2=Q_G^2=0$.
These identities give quite powerful constraints to determine the equations of motion
in the next section.

\section{Equations of motion including the R sector}\label{eom including R sector}

The full equations of motion for the NS and the R sectors 
of the heterotic string field theory are 
non-polynomial not only in the NS string field $V$ but also 
in the R string field $\Psi$. In this section we find their explicit forms order 
by order in $\Psi$ by imposing some consistency conditions.
We also determine a subset of the terms to all orders.

We start with the equation of motion for the NS sector\footnote{
A factor $-1/2$ in the second term, or equivalently the normalization of $\Psi$,
is chosen in such a way that fermion amplitudes are reproduced with the correct
normalization as shown in the next section.}
(the NS equation of motion),
\begin{equation}
 \eta  G-\frac{\kappa}{2}\mspd{\epsi^2}=0,\label{leading eom NS}
\end{equation}
including the leading-order coupling to the R string $\Psi$.
This is a straightforward extension of that of the open superstring 
field theory.\cite{Berkovits:2001im}
We use here the string product shifted by $G(V)$,
on which the shifted BRST operator $Q_G$ acts as derivation.
As a result, (\ref{leading eom NS}) becomes consistent with 
the identity (\ref{id1}) by taking
\begin{equation}
  Q_G\eta\Psi=0\label{leading eom R}
\end{equation}
to be the equation of motion for the R sector (the R equation of motion), 
a non-linear extension of 
(\ref{linearized eom R}). In contrast, (\ref{leading eom NS}) 
is not consistent with another identity (\ref{id2}) because $\eta$ fails to
be derivation of the shifted string products.
If we apply $\eta$ on the left-hand side of (\ref{leading eom NS}), we obtain
\begin{equation}
 \eta\left(\eta G -\frac{\kappa}{2}\mspd{\epsi^2}\right) =
\frac{\kappa^2}{2}\mspd{\eta G,\epsi^2},\label{inconsistent}
\end{equation}
where the right-hand side is non-zero under (\ref{leading eom NS}) although
it is higher order in $\Psi$.
This inconsistency can be improved order by order in $\Psi$
by adding the correction terms to the equations of motion.

Before looking for the next order corrections, 
we consider here the gauge transformations.
We first note that (\ref{leading eom NS}) and (\ref{leading eom R}) 
are invariant under the $\Lambda_0$-gauge transformation, 
\begin{equation}
 \bar{B}_{\delta_{\Lambda_0}}(V) = Q_G\Lambda_0,\qquad 
\delta_{\Lambda_0}\Psi = 0,\label{Lambda zero}
\end{equation}
since it keeps $G(V)$ invariant:
\begin{equation}
 \delta_{\Lambda_0} G(V) = 
Q_G\bar{B}_{\delta_{\Lambda_0}}(V) = Q_G(Q_G\Lambda_0) = 0.
\end{equation}

From the leading order $\Lambda_1$-gauge transformation of $V$, (\ref{NS gauge tf}), 
we can determine that of $\Psi$ so that the R equation of motion 
(\ref{leading eom R}) is invariant up to higher-order terms in $\Psi$.
The leading-order gauge transformation is, in total, given by
\begin{equation}
 \bar{B}_{\delta^{(0)}_{\Lambda_1}}(V)=\eta\Lambda_1,\qquad
\delta^{(0)}_{\Lambda_1}\Psi=-\kappa\mspd{\Psi,\eLambda},\label{leading lambda one tf}
\end{equation}
where we explicitly indicated, as the superscript on $\delta$,
the fermion number difference between before and after the transformation.
In fact, the R equation of motion
(\ref{leading eom R}) is transformed under this transformation as
\begin{equation}
  \delta^{(0)}_{\Lambda_1}(Q_G\eta\Psi) =
-\kappa\mspd{Q_G\eta\Psi,\eLambda}
+\kappa^2Q_G\left(\mspd{\eta G,\Psi,\eLambda}\right),
\end{equation}
where the second term is $\mathcal{O}(\Psi^3)$ under (\ref{leading eom NS}).
The next order correction,
 \begin{equation}
 \bar{B}_{\delta^{(2)}_{\Lambda_1}}(V)=-\frac{\kappa^2}{2}\mspd{\Psi,\eta\Psi,\eLambda},
\end{equation}
has to be included in the transformation of $V$
to keep the NS equation of motion (\ref{leading eom NS}) invariant,
except for the $\mathcal{O}(\Psi^4)$ terms, as
\begin{align}
(\delta^{(0)}_{\Lambda_1}+\delta^{(2)}_{\Lambda_1})\left(\eta
 G-\frac{\kappa}{2}\mspd{\epsi^2}\right) =&
-\kappa\mspd{\left(\eta G-\frac{\kappa}{2}\mspd{\epsi^2}\right),\eLambda}
+\kappa^2\mspd{Q_G\eta\Psi,\eta\Psi,\eLambda}\nonumber\\
&+\frac{\kappa^3}{2}\mspd{\eta G,\mspd{\Psi,\eta\Psi,\eLambda}}
-Q_G\left(\frac{\kappa^3}{2}\mspd{\eta G,\Psi,\eta\Psi,\eLambda}\right)\nonumber\\
&-\kappa^3\mspd{\eta\Psi,\mspd{\eta G,\Psi,\eLambda}}+\mathcal{O}(\Psi^4).
\end{align}

The leading-order $\Lambda_{\frac{1}{2}}$-gauge transformation can similarly be determined as
\begin{equation}
 \bar{B}_{\delta^{(0)}_{\Lambda_{1/2}}}(V)=0,\qquad
\delta^{(0)}_{\Lambda_{1/2}}\Psi=Q_G\Lambda_{\frac{1}{2}},\label{leading lambda one-half tf} 
\end{equation}
so that the R equation of motion (\ref{leading eom R}) is invariant:
\begin{equation}
 \delta^{(0)}_{\Lambda_{1/2}}(Q_G\eta\Psi) = -\kappa\mspd{\eta G,\qLambda}.
\end{equation}
The right-hand side is equal to zero except for $\mathcal{O}(\Psi^3)$ terms\footnote{
We denote the terms with $n$ R string fields, including
not only $\Psi$ but also $\Lambda_{1/2}$, $\Lambda_{3/2}$, and $\Xi$,
by $\mathcal{O}(\Psi^n)$ in this paper.}
under (\ref{leading eom NS}). The invariance of the NS equation of motion
(\ref{leading eom NS}) requires the next order correction
\begin{equation}
 \bar{B}_{\delta^{(2)}_{\Lambda_{1/2}}}(V)=\kappa\mspd{\eta\Psi,\Lambda_{\frac{1}{2}}}.
\end{equation}
Then,
\begin{align}
(\delta^{(0)}_{\Lambda_{1/2}}+\delta^{(2)}_{\Lambda_{1/2}})
\left(\eta G-\frac{\kappa}{2}\mspd{\epsi^2}\right) =&
\kappa^2\mspd{\left(\eta G-\frac{\kappa}{2}\mspd{\epsi^2}\right),\eta\Psi,\qLambda}\nonumber\\
&-\eta\left(\kappa\mspd{Q_G\eta\Psi,\Lambda_{\frac{1}{2}}}\right)
+\mathcal{O}(\Psi^4).
\end{align}

The invariance under the leading-order $\Lambda_{\frac{3}{2}}$-gauge transformation,
\begin{equation}
 \bar{B}_{\delta^{(2)}_{\Lambda_{3/2}}}(V)=0,\qquad \delta^{(0)}_{\Lambda_{3/2}}\Psi=\eLambdaf,
\end{equation}
is exact in this order
since $\Psi$ appears only in the form of $\eta\Psi$ so far.

In order to consider further corrections, we assume that 
the $\Lambda_0$-gauge transformation (\ref{Lambda zero}) receives
no more correction and the NS string field $V$ only appears 
in the correction terms through $G(V)$ in the quantities $Q_G$ and $\mspd{\cdots}$, 
which keeps the equations of motion invariant under 
(\ref{Lambda zero}). We call this assumption \textit{the G-ansatz} in this paper.
Then the consistency with the identity (\ref{id1})
requires the full equations of motion to be in the form of
\begin{subequations}\label{EOM}
 \begin{align}
 \eta G - \frac{\kappa}{2}\mspd{\Omega^2}+Q_G\Sigma =& 0,\label{NS eom}\\
 Q_G\Omega =& 0,\label{R eom}
\end{align}
\end{subequations}
as explained in detail in Appendix \ref{appendix B},
where the string fields $\Omega$ and the $\Sigma$ have $(G,P)=(2,-1/2)$ and $(2,-1)$,
respectively, and can be expanded in $\Psi$ as
\begin{equation}
 \Omega = \sum_{n=0}^\infty \Omega^{(2n+1)},\qquad 
 \Sigma = \sum_{n=0}^\infty \Sigma^{(2n+2)},
\end{equation}
with the leading terms 
\begin{equation}
\Omega^{(1)}=\eta\Psi,\qquad \Sigma^{(2)}=0.\label{L omega and sigma}
\end{equation}
The number in the parentheses of the superscript denotes the fermion number,
that is the number of $\Psi$, of the terms included.
%
%
One can see, by counting the ghost and the picture numbers using (\ref{g and p}), 
that the fundamental possible terms in the $\Omega^{(2n+1)}$ and the $\Sigma^{(2n+2)}$ 
at each order $n\ (\ge1)$ take the form,
\begin{equation}
\kappa^{2n+\alpha}\mspd{\Psi,\qpsi^{k_1},\epsi^{l_1},\mspd{\Psi,\qpsi^{k_2},\epsi^{l_2},
\mspd{\cdots,\mspd{\Psi,\qpsi^{k_m},\epsi^{l_m}}\cdots}}},\label{possible term}
\end{equation}
under the G-ansatz, with $\alpha=0$ $(1)$ for $\Omega^{(2n+1)}$ $(\Sigma^{(2n+2)})$.
Each term is made by use of $m$ ($1\le m\le n$) string products\footnote{
Hereafter we refer to the shifted quantities, $Q_G$ and $\mspd{\cdots}$, 
as simply the BRST operator and the string product, respectively.}
with the non-negative integers $k_i$ and $l_i$ $(1\le i\le m)$ 
restricted by the equations, 
$\sum_{i=1}^mk_i=n-m$, $\sum_{i=1}^ml_i=n+1+\alpha$ and $k_m+l_m\ne0$. 
More general possible terms, appearing for ($n\ge3$), can be obtained 
by replacing the $\eta\Psi$s in these fundamental terms with fundamental terms 
in $\Omega$, such as
\begin{align}
\kappa^{2n+\alpha}\mspd{\Psi,&\qpsi^{k_1},\epsi^{l_1-3},\Omega_1^{(2n_1+1)},(\Omega_2^{(2n_2+1)})^2,
\nonumber\\
&\times\mspd{\Psi,\qpsi^{k_2},\epsi^{l_2-1},\Omega_3^{(2n_3+1)},
\mspd{\cdots,\mspd{\Psi,\qpsi^{k_m},\epsi^{l_m}}\cdots}}},\label{general possible term}
\end{align}
with $\sum_{i=1}^mk_i=n-n_1-2n_2-n_3-m$, $\sum_{i=1}^ml_i=n-n_1-2n_2-n_3+1+\alpha$ and
$k_m+l_m\ne0$, where
\begin{align}
 \Omega_1^{(2n_1+1)}=&\kappa^{2n_1}\mspd{\Psi,\qpsi^{k_{11}},\epsi^{l_{11}},\mspd{\Psi,\qpsi^{k_{12}},\epsi^{l_{12}},
\nonumber\\
&\hspace{3cm}
\times
\mspd{\cdots,\mspd{\Psi,\qpsi^{k_{1m_1}},\epsi^{l_{1m_1}}}\cdots}}},\\
 \Omega_2^{(2n_2+1)}=&\kappa^{2n_2}\mspd{\Psi,\qpsi^{k_{21}},\epsi^{l_{21}},\mspd{\Psi,\qpsi^{k_{22}},\epsi^{l_{22}},
\nonumber\\
&\hspace{3cm}
\times\mspd{\cdots,\mspd{\Psi,\qpsi^{k_{2m_2}},\epsi^{l_{2m_2}}}\cdots}}},\\
 \Omega_3^{(2n_3+1)}=&\kappa^{2n_3}\mspd{\Psi,\qpsi^{k_{31}},\epsi^{l_{31}},\mspd{\Psi,\qpsi^{k_{32}},\epsi^{l_{32}},
\nonumber\\
&\hspace{3cm}
\times\mspd{\cdots,\mspd{\Psi,\qpsi^{k_{3m_3}},\epsi^{l_{3m_3}}}\cdots}}},
\end{align}
with $\sum_{j=1}^{m_i}k_{ij}=n_i-m_i$, $\sum_{j=1}^{m_i}l_{ij}=n_i+1$ and
$k_{im}+l_{im}\ne0$ for $i=1,2,3$. Moreover, the terms obtained by moving the positions of $\eta\Psi$s 
are also allowed. For example, the terms 
\begin{equation} 
\mspd{\Psi,\eta\Psi,\mspd{\Psi,\eta\Psi},\mspd{\Psi,\epsi^2}},\quad
\mspd{\Psi,\mspd{\Psi,\eta\Psi},\mspd{\Psi,\epsi^3}},\quad
\mspd{\Psi,\epsi^2,\mspd{\Psi,\eta\Psi},\mspd{\Psi,\eta\Psi}}
\end{equation}
are also possible in $\Omega^{(7)}$, which are obtained 
by moving $\eta\Psi$s in a term
\begin{equation}
\mspd{\Psi,(\Omega^{(3)})^2} \sim \mspd{\Psi,\mspd{\Psi,\epsi^2},\mspd{\Psi,\epsi^2}}.  
\end{equation}
After these replacements and  moves, 
we can again replace the $\eta\Psi$s with fundamental terms in $\Omega$ and move
the $\eta\Psi$s for higher $n$. All the possible terms are obtained by repeating
these replacements and moves as many times as possible for given $n$. 

From these general considerations, we can find that 
the possible next-to-leading-order corrections to the equations
of motion are given by
\begin{equation}
 \Omega^{(3)} = -\frac{\kappa^2}{3!}\mspd{\Psi,\epsi^2},\qquad
 \Sigma^{(4)} = \frac{\kappa^3}{4!}\mspd{\Psi,\epsi^3},\label{NL omega and sigma}
\end{equation}
where the numerical coefficients were fixed by solving the consistency equation,
\begin{equation}
 \eta\left(\frac{\kappa}{2}\mspd{\Omega^2}-Q_G\Sigma\right) = 0,\label{consistency id2}
\end{equation}
with the identity (\ref{id2}).

As in the case of the leading order, we can obtain the gauge transformations
in the next-to-leading order, $\delta^{(2)}\Psi$ and $\bar{B}_{\delta^{(4)}}$,
by imposing the invariance of the R and the NS equations of motion, respectively.
The results are
\begin{align}
\bar{B}_{\delta^{(4)}_{\Lambda_1}}(V) =&
-\frac{\kappa^4}{4!}\mspd{\Psi,Q_G\Psi,\epsi^2,\eLambda}
-\frac{\kappa^4}{8}\mspd{\mspd{\Psi,\eta\Psi},\Psi,\eta\Psi,\eLambda}\nonumber\\
&+\frac{\kappa^4}{4!}\mspd{\mspd{\Psi,\epsi^2},\Psi,\eLambda}
+\frac{\kappa^4}{4!}\mspd{\Psi,\mspd{\Psi,\epsi^2,\eLambda}}\nonumber\\
&+\frac{\kappa^4}{8}\mspd{\Psi,\eta\Psi,\mspd{\Psi,\eta\Psi,\eLambda}},\nonumber\\
 \delta^{(2)}_{\Lambda_1}\Psi =&
-\frac{\kappa^3}{3!}\mspd{\Psi,Q_G\Psi,\eta\Psi,\eLambda}
-\frac{\kappa^3}{3}\mspd{\mspd{\Psi,\eta\Psi},\Psi,\eLambda}\nonumber\\
&+\frac{\kappa^3}{3!}\mspd{\Psi,\mspd{\Psi,\eta\Psi,\eLambda}},
\end{align}
for $\Lambda_1$-transformation,
\begin{align}
\bar{B}_{\delta^{(4)}_{\Lambda_{1/2}}}(V) =&
\frac{\kappa^3}{8}\mspd{\Psi,\epsi^2,\qLambda}
-\frac{\kappa^3}{3!}\mspd{\mspd{\Psi,\epsi^2},\Lambda_{\frac{1}{2}}},\nonumber\\
\delta^{(2)}_{\Lambda_{1/2}}\Psi =& \frac{\kappa^2}{3}\mspd{\Psi,\eta\Psi,\qLambda},
\end{align}
for $\Lambda_{1/2}$-transformation and
\begin{equation}
 \bar{B}_{\delta^{(4)}_{\Lambda_{3/2}}}(V) = -\frac{\kappa^3}{4!}\mspd{\Psi,\epsi^2,\eLambdaf},\qquad
\delta^{(2)}_{\Lambda_{3/2}}\Psi =
- \frac{\kappa^2}{3!}\mspd{\Psi,\eta\Psi,\eLambdaf},
\end{equation}
for $\Lambda_{3/2}$-transformation.

The next-to-next-to-leading corrections the $\Omega^{(5)}$ and the $\Sigma^{(6)}$ 
are further determined by solving (\ref{consistency id2}) as
\begin{align}
 \Omega^{(5)} = &
-\frac{\kappa^4}{5!}\mspd{\Psi,Q_G\Psi,\epsi^3}
+\frac{\kappa^4}{5!}\mspd{\Psi,\mspd{\Psi,\epsi^3}}
+\frac{4}{5!}\kappa^4\mspd{\Psi,\eta\Psi,\mspd{\Psi,\epsi^2}}\nonumber\\
&-\frac{4}{5!}\kappa^4\mspd{\Psi,\epsi^2,\mspd{\Psi,\eta\Psi}},\nonumber\\
\Sigma^{(6)} =& 
\frac{\kappa^5}{6!}\mspd{\Psi,Q_G\Psi,\epsi^4}
-\frac{\kappa^5}{6!}\mspd{\Psi,\mspd{\Psi,\epsi^4}}
-\frac{5}{6!}\kappa^5\mspd{\Psi,\eta\Psi,\mspd{\Psi,\epsi^3}}\nonumber\\
&-\frac{10}{6!}\kappa^5\mspd{\Psi,\epsi^2,\mspd{\Psi,\epsi^2}}
+\frac{5}{6!}\kappa^5\mspd{\Psi,\epsi^3,\mspd{\Psi,\eta\Psi}}.\label{NNL omega and sigma}
\end{align}
The gauge transformations $\bar{B}_{\delta^{(6)}}$ and $\delta^{(4)}\Psi$
are also obtained similarly and their explicit forms are given 
in Appendix \ref{appendix A} since they are lengthy.

In general, we conjecture that the full equations of motion can 
be constructed order by order in $\Psi$ in a similar manner. In order to show that 
this is actually the case, we have to prove the existence of 
a non-trivial, but not necessarily unique, solution to the consistency equation
(\ref{consistency id2}). If it has a unique solution, we can, in principle, 
obtained the full equations of motion and also the gauge transformations 
in any order in $\Psi$ in the same way as above.
If a non-trivial solution exists but is not unique, we must take into account 
the further gauge invariance(s) to find the full equations of motion, 
where the equations of motion and the gauge transformation(s) have to be 
determined simultaneously.

While we have not explicitly found all the terms of the full equations of motion, 
we have completely determined a subset of the terms
built with one string product in the $\Omega$ and the $\Sigma$ and those with two string
products in the $\Sigma$ to all orders in $\Psi$ similarly by solving (\ref{consistency id2}). 
The result is \begin{subequations}\label{sp expansion}
  \begin{align}
 \Omega =& \eta\Psi\ -\ 
\sum_{n=1}^\infty\frac{\kappa^{2n}}{(2n+1)!}\mspd{\Psi,\qpsi^{n-1},\epsi^{n+1}}+\cdots,
\label{sp expansion in omega}\\
\Sigma =&\sum_{n=1}^\infty\frac{\kappa^{2n+1}}{(2n+2)!}\mspd{\Psi,\qpsi^{n-1},\epsi^{n+2}}
\nonumber\\
&+\sum_{n=2}^\infty
\underset{(m,l)\ne(0,0)}{\sum_{m=0}^{n-2}\sum_{l=0}^{n+2}}
g_{n,m,l}\kappa^{2n+1}\mspd{\Psi,\qpsi^{n-m-2},\epsi^{n-l+2},\mspd{\Psi,\qpsi^m,\epsi^l}}+\cdots,
\label{sp expansion in sigma}
\end{align}
\end{subequations}
with
\begin{equation}
g_{n,m,l}=
\begin{cases}
\displaystyle
\sum_{k=0}^l\comb{n-1}{m+l+1-k}\comb{n+3}{k}\frac{1}{(2n+2)!} & \textrm{for}\ (0\le l\le m+1), \\
\displaystyle
-\sum_{k=0}^m\comb{n-1}{k}\comb{n+3}{m+l+1-k}\frac{1}{(2n+2)!} & \textrm{for}\  (m+2\le l\le n+2).
\end{cases}
\end{equation}
The dots in the right-hand side of (\ref{sp expansion}) represent the terms built with more string products
and $\smallcomb{n}{m}$ is the binomial coefficient defined by
\begin{equation}
 \comb{n}{m}=
\begin{cases}
\displaystyle \frac{n!}{(n-m)!m!} & \textrm{for}\ (0\le m\le n),\\
\ 0 & \textrm{for}\ (m<0,\ \textrm{or}\ \ m>n).
\end{cases}
\end{equation}
The derivation of (\ref{sp expansion}) is presented in more detail 
in Appendix \ref{appendix C}. 
The gauge transformations in the corresponding approximation
are also found in a similar way to those in the fermion expansions as
\begin{align}
&\bar{B}_{\delta_{\Lambda_1}}(V) = \eta\Lambda_1
-\sum_{n=1}^\infty\frac{\kappa^{2n}}{(2n)!}\mspd{\Psi,\qpsi^{n-1},\epsi^n,\eLambda}
+\bar{B}^{[2]}_{\delta_{\Lambda_1}}(V)+\cdots,\nonumber\\
&\delta_{\Lambda_1}\Psi =
 -\sum_{n=0}^\infty\frac{\kappa^{2n+1}}{(2n+1)!}\mspd{\Psi,\qpsi^n,\epsi^n,\eLambda}
+\cdots,\\
&\bar{B}_{\delta_{\Lambda_{1/2}}}(V) = \kappa\mspd{\eta\Psi,\Lambda_{\frac{1}{2}}}
+\sum_{n=1}^\infty\frac{(n+2)\kappa^{2n+1}}{(2n+2)!}\mspd{\Psi,\qpsi^{n-1},\epsi^{n+1},\qLambda}
+\bar{B}^{[2]}_{\delta_{\Lambda_1}}(V)+\cdots,\nonumber\\
&\delta_{\Lambda_{1/2}}\Psi = Q_G\Lambda_{\frac{1}{2}}
+\sum_{n=1}^\infty\frac{(n+1)\kappa^{2n}}{(2n+1)!}\mspd{\Psi,\qpsi^{n-1},\epsi^n,\qLambda}+\cdots,\\
&\bar{B}_{\delta_{\Lambda_{3/2}}}(V) =
-\sum_{n=2}^\infty\frac{(n-1)\kappa^{2n-1}}{(2n)!}\mspd{\Psi,\qpsi^{n-2},\epsi^n,\eLambdaf}
+\bar{B}^{[2]}_{\delta_{\Lambda_1}}(V)+\cdots,\nonumber\\
&\delta_{\Lambda_{3/2}}\Psi = \eta\Lambda_{\frac{3}{2}}
-\sum_{n=1}^\infty\frac{n\kappa^{2n}}{(2n+1)!}\mspd{\Psi,\qpsi^{n-1},\epsi^n,\eLambdaf}+\cdots,
\end{align}
where the terms built with two string products are denoted by $\bar{B}^{[2]}_{\delta}(V)$'s and 
explicitly given in Appendix \ref{appendix C}.
We regard these results as evidence that the consistency equation 
(\ref{consistency id2}) actually has a non-trivial solution
and we can construct the full equations of motion and gauge transformations 
to all orders in $\Psi$.

\section{Action with constraint}\label{action with a constraint}

As in the case of the open superstring field theory,\cite{Michishita:2004by} 
one can construct an action by introducing an auxiliary R string field $\Xi$
with a constraint. The above equations of motion can be obtained from 
those derived from this action after eliminating $\Xi$ using 
the constraint.

Let us start with the bilinear action
\begin{equation}
S_R^{(2)} = \frac{1}{2}\langle \eta\Psi, Q_G\Xi \rangle,\label{R action 2}
\end{equation} 
which is a natural extension of the action in the open superstring field 
theory.\cite{Michishita:2004by}\footnote{
See also \S\S\ref{Feynman rule}.} In order for this action (\ref{R action 2}) 
to be non-zero, the auxiliary string field $\Xi$ must have the properties:
\begin{equation}
 \Xi\ :\ \textrm{Grassmann\ odd},\ (G,P)=(1,-\frac{1}{2}).
\end{equation}
From the action $S_{NS}+S_R^{(2)}$, we can derive the 
three equations of motion,
\begin{equation}
  \eta G - \frac{\kappa}{2}\mspd{\eta\Psi,Q_G\Xi} = 0,\qquad
 Q_G\eta\Psi = 0,\qquad
 \eta Q_G\Xi = 0,
\end{equation}
which are equivalent to the equations of motion with the leading-order coupling,
(\ref{leading eom NS}) and (\ref{leading eom R}), if we impose
the constraint, $Q_G\Xi=\eta\Psi$.

In general, the full R action is given by the infinite series
\begin{equation}
 S_R = \sum_{n=1}^{\infty}S_R^{(2n)},\label{R action} 
\end{equation}
where $S_R^{(2n)}$ includes $2n$ R string fields, the $\Psi$ and the $\Xi$, 
in total. If we take the first three terms as (\ref{R action 2}) and
\begin{align}
S_R^{(4)} =& 
\frac{\kappa^2}{4!}\langle \eta\Psi, \mspd{\Psi,\qxi^2}\rangle,\label{R action 4}\\
S_R^{(6)} =& \frac{\kappa^4}{6!}\langle\eta\Psi,\mspd{\Psi,\qpsi,\qxi^3}\rangle
 -\frac{2}{6!}\kappa^4\langle\eta\Psi, \mspd{\Psi,\mspd{\Psi,\qxi^3}} \rangle\nonumber\\
 & +\frac{2}{6!}\kappa^4\langle\eta\Psi, \mspd{\Psi,\qxi,\mspd{\Psi,\qxi^2}} \rangle
 +\frac{3}{6!}\kappa^4\langle\eta\Psi, \mspd{\Psi,\qxi^2,\mspd{\Psi,\qxi}} \rangle,
\label{R action 6}
\end{align}
the action $S_{NS}+S_R^{(2)}+S_R^{(4)}+S_R^{(6)}$
yields the equations of motion,
\begin{equation}
 \eta G - \frac{\kappa}{2}\mspd{\tilde{\Omega},\qxi}+Q_G\tilde{\Sigma}=0,\qquad
 Q_G\tilde{\Omega}=0,\qquad
 \eta \Theta + \Delta=0,\label{NNL eom from action}
\end{equation}
with
\begin{align}
 \tilde{\Omega}=&\eta\Psi - \frac{\kappa^2}{3!}\mspd{\Psi,\eta\Psi,\qxi}\nonumber\\
&-\frac{\kappa^4}{5!}\mspd{\Psi,\qpsi,\eta\Psi,\qxi^2}
+\frac{\kappa^4}{5!}\mspd{\Psi,\mspd{\Psi,\eta\Psi,\qxi^2}}\nonumber\\
&-\frac{2}{3\cdot5!}\kappa^4\mspd{\Psi,\eta\Psi,\mspd{\Psi,\qxi^2}}
+\frac{4}{3\cdot5!}\kappa^4\mspd{\Psi,\qxi,\mspd{\Psi,\eta\Psi,\qxi}}\nonumber\\
&-\frac{2}{5!}\kappa^4\mspd{\Psi,\eta\Psi,\qxi,\mspd{\Psi,\qxi}}
-\frac{2}{5!}\kappa^4\mspd{\Psi,\qxi^2,\mspd{\Psi,\eta\Psi}},\\
 \tilde{\Sigma}=&\frac{\kappa^3}{4!}\mspd{\Psi,\eta\Psi,\qxi^2}\nonumber\\
&+\frac{\kappa^5}{6!}\mspd{\Psi,\qpsi,\eta\Psi,\qxi^3}
-\frac{\kappa^5}{6!}\mspd{\Psi,\mspd{\Psi,\eta\Psi,\qxi^3}}\nonumber\\
&-\frac{2}{6!}\kappa^5\mspd{\Psi,\eta\Psi,\mspd{\Psi,\qxi^3}}
-\frac{3}{6!}\kappa^5\mspd{\Psi,\qxi,\mspd{\Psi,\eta\Psi,\qxi^2}}\nonumber\\
&+\frac{2}{6!}\kappa^5\mspd{\Psi,\eta\Psi,\qxi,\mspd{\Psi,\qxi^2}}
-\frac{2}{6!}\kappa^5\mspd{\Psi,\qxi^2,\mspd{\Psi,\eta\Psi,\qxi}}\nonumber\\
&+\frac{3}{6!}\kappa^5\mspd{\Psi,\eta\Psi,\qxi^2,\mspd{\Psi,\qxi}}
+\frac{2}{6!}\kappa^5\mspd{\Psi,\qxi^3,\mspd{\Psi,\eta\Psi}},\\
 \Theta=&Q_G\Xi+\frac{\kappa^2}{3!}\mspd{\Psi,\qxi^2}\nonumber\\
&+\frac{\kappa^4}{5!}\mspd{\Psi,\qpsi,\qxi^3}
-\frac{\kappa^4}{5!}\mspd{\Psi,\mspd{\Psi,\qxi^3}}\nonumber\\
&+\frac{\kappa^4}{45}\mspd{\Psi,\qxi,\mspd{\Psi,\qxi^2}}
+\frac{4}{5!}\kappa^4\mspd{\Psi,\qxi^2,\mspd{\Psi,\qxi}},\\
\Delta=&\frac{\kappa^4}{3\cdot5!}\mspd{\eta\Psi,\qxi,\mspd{\Psi,\qxi^2}}
-\frac{\kappa^4}{3\cdot5!}\mspd{\qxi^2,\mspd{\Psi,\eta\Psi,\qxi}}\nonumber\\
&+\frac{\kappa^4}{3\cdot5!}\mspd{\Psi,\eta\Psi,\mspd{\qxi^3}}
-\frac{\kappa^4}{3\cdot5!}\mspd{\Psi,\qxi,\mspd{\eta\Psi,\qxi^2}}\nonumber\\
&+\frac{\kappa^4}{5!}\mspd{\eta\Psi,\qxi^2,\mspd{\Psi,\qxi}}
-\frac{\kappa^4}{5!}\mspd{\qxi^3,\mspd{\Psi,\eta\Psi}}\nonumber\\
&+\frac{\kappa^4}{5!}\mspd{\Psi,\eta\Psi,\qxi,\mspd{\qxi^2}}
-\frac{\kappa^4}{5!}\mspd{\Psi,\qxi^2,\mspd{\eta\Psi,\qxi}}.
\label{eom from action}
\end{align}
One can show that if we impose the constraint 
\begin{equation}
 Q_G\Xi = \Omega,\label{constraint}
\end{equation} 
these quantities become 
\begin{equation}
 \tilde{\Omega}=\Omega,\qquad
 \tilde{\Sigma}=\Sigma,\qquad
 \Theta=\eta\Psi,\qquad
 \Delta=0,
\end{equation}
with (\ref{L omega and sigma}), (\ref{NL omega and sigma}), and (\ref{NNL omega and sigma}),
and thus the equations of motion (\ref{NNL eom from action}) reduce to 
those obtained in the previous section. 

Although we can also consider the gauge transformations, the action 
$S_{NS}+S_R^{(2)}+S_R^{(4)}+S_R^{(6)}$ has less symmetry than 
the corresponding equations of motion written by the $V$ and the $\Psi$.
The action is only invariant under the transformations,
\begin{equation}
 \bar{B}_{\delta_{\Lambda_0}}(V) = Q_G\Lambda_0,\qquad \delta\Psi=0,\qquad
  \delta\Xi=Q_G\Lambda_{-\frac{1}{2}}.\label{action symmetry}
\end{equation}
This symmetry enhancement (decline) also happens in the open superstring field 
theory,\cite{Michishita:2004by} but the action symmetry (\ref{action symmetry}) is smaller. 
The gauge symmetry generated by $\Lambda_{-1/2}$ requires that the auxiliary field $\Xi$ 
have to take the form of $Q_G\Xi$ in the action.

We have also completely determined a subset of terms, 
built with the one string product, in the full action (\ref{R action}) 
to all orders in $\Psi$ as
\begin{equation}
 S_R = \frac{1}{2}\langle\eta\Psi,Q_G\Xi\rangle+
\sum_{n=1}^\infty \frac{\kappa^{2n}}{(2n+2)!}
\langle\eta\Psi,\mspd{\Psi,\qpsi^{n-1},\qxi^{n+1}}\rangle + \cdots.
\end{equation}
This reproduces the terms with one string product in (\ref{sp expansion})
by imposing the constraint (\ref{constraint}) with (\ref{sp expansion in omega}).

\section{Feynman rules and four--point amplitudes}\label{Feynman rules}

We propose in this section the Feynman rules to compute tree-level 
amplitudes in the heterotic string field theory
by extending those for the open superstring.\cite{Michishita:2004by} 
We show that these rules actually reproduce the correct on-shell four-point
amplitudes with external fermions.

\subsection{Feynman rules for tree-level amplitudes}\label{Feynman rule}

Let us first focus on the kinetic terms 
\begin{equation}
S_0=\frac{1}{2}\langle\eta V,Q V\rangle
+\frac{1}{2}\langle\eta\Psi,Q\Xi\rangle,\label{free action}
\end{equation}
in the action, which are invariant under the transformations,
\begin{equation}
\delta V=Q\Lambda_0+\eta\Lambda_1,\qquad
\delta\Psi=Q\Lambda_{\frac{1}{2}}+\eta\Lambda_{\frac{3}{2}},\qquad
\delta\Xi=Q\Lambda_{-\frac{1}{2}}+\eta\Lambda'_{\frac{1}{2}}.\label{free action symmetry}
\end{equation}
The non-trivial point is that not all of these symmetries can be extended to those of
the full action. In particular, it may be worthwhile to note that 
even the NS propagator cannot be derived by means of the conventional method
in the heterotic string field theory since the action symmetry (\ref{action symmetry})
does not include that generated by $\Lambda_1$.

Here we simply assume that we can \lq\lq effectively\rq\rq\ 
fix these symmetries (\ref{free action symmetry}) by the gauge conditions
\begin{equation}
  b_0^+V=\xi_0V=0,\qquad
   b_0^+\Psi=\xi_0\Psi=0,\qquad
  b_0^+\Xi=\xi_0\Xi=0.
\label{gauge conditions}
\end{equation}
Then, by inverting the kinetic terms (\ref{free action}),
the propagators in this gauge are given by
\begin{align}
\overbracket[0.5pt]{\!\!VV\!\!} \equiv& \Pi_{NS} = \xi_0\frac{b_0^+b_0^-}{L_0^+}\delta(L_0^-)
=\xi_0b_0^+b_0^-\int_0^\infty \!\!dT\!\!\int_0^{2\pi}\frac{d\theta}{2\pi}e^{-TL^+_0-i\theta L^-_0},
\label{NS propagator}\\
 \overbracket[0.5pt]{\!\!\Psi\Xi\!\!} = \overbracket[0.5pt]{\!\!\Xi\Psi\!\!}
\equiv& \Pi_R = 2\xi_0\frac{b_0^+b_0^-}{L_0^+}\delta(L_0^-)=2\Pi_{NS}.\label{R propagator}
\end{align}

As in the open superstring field theory,\cite{Michishita:2004by}
the constraint can be taken into account by replacing the $Q\Xi$  
and the $\eta\Psi$ with their \textit{self-dual} part $\omega=(Q\Xi+\eta\Psi)/2$
in the vertices. However, the $\Xi$ and the $\Psi$ do not always appear in the action
(\ref{R action})
in the form of the $Q\Xi$ and the $\eta\Psi$, respectively.
Some preparation is needed in advance of the replacement.

First, we point out that the open superstring action can be written 
in a similar form of the heterotic string action by using the relation,
\begin{equation}
Q'A =QA +[(e^{-\Phi}Qe^\Phi),A\} = e^{-\Phi}(Q\tilde{A})e^\Phi,\label{oss relation}
\end{equation}
with $\tilde{A}=e^\Phi A e^{-\Phi}$. 
It is reasonable to expect that
a similar relation enables us to rewrite the action (\ref{R action}) 
in the opposite direction so that the redefined auxiliary field $\tilde{\Xi}$ 
always appears in the form of $Q\tilde{\Xi}$. 
Actually, we can show, up to $\mathcal{O}(\kappa^3)$, that the relation
\begin{align}
 Q_G\Xi =& Q\tilde{\Xi}+\kappa[V,Q\tilde{\Xi}]+\frac{\kappa^2}{2}[V,QV,Q\tilde{\Xi}]
+\frac{\kappa^2}{2}[V,[V,Q\tilde{\Xi}]]
\nonumber\\
&
+\frac{\kappa^3}{3!}[V,(QV)^2,Q\tilde{\Xi}]
+\frac{\kappa^3}{3!}[V,[V,QV,Q\tilde{\Xi}]]
+\frac{\kappa^3}{3}[V,QV,[V,Q\tilde{\Xi}]]
\nonumber\\
&
+\frac{\kappa^3}{3!}[[V,QV],V,Q\tilde{\Xi}]
+\frac{\kappa^3}{3!}[V,[V,[V,Q\tilde{\Xi}]]]+\cdots,\label{redefinition}
\end{align}
holds, where
\begin{align}
\tilde{\Xi} =& \Xi-\kappa[V,\Xi]-\frac{\kappa^2}{2}[V,QV,\Xi]+\frac{\kappa^2}{2}[V,[V,\Xi]]
-\frac{\kappa^3}{3!}[V,(QV)^2,\Xi]+\frac{\kappa^3}{3}[V,[V,QV,\Xi]]\nonumber\\
&+\frac{\kappa^3}{3!}[V,QV,[V,\Xi]]-\frac{\kappa^3}{3!}[[V,QV],V,\Xi]-\frac{\kappa^3}{3!}[V,[V,[V,\Xi]]]
+\cdots.\label{point tf}
\end{align}
If we also note that $\Psi=\xi_0\epsi$ under the gauge conditions 
(\ref{gauge conditions}),
the action (\ref{R action}) can be rewritten in such a way that 
all the $\tilde{\Xi}$ and the $\Psi$ appear in the form of $Q\tilde{\Xi}$ and $\eta\Psi$, 
respectively.
As a consequence, we can completely project out the component which does not satisfy 
the (linearized) constraint $Q\tilde{\Xi}=\eta\Psi$ by replacing 
$Q\tilde{\Xi}$ and $\eta\Psi$ with $\omega=(Q\tilde{\Xi}+\eta\Psi)/2$.

In the end of this subsection, we must mention that the prescription proposed here can
give definite rules but has an ambiguity coming from the fact that $\eta\Psi=\eta\xi_0\epsi$ 
but $\omega\ne\eta\xi_0\omega$. 
We take here a convention that $\eta\Psi$, $\Psi$, and $Q\Psi$ are replaced 
with $\omega$, $\xi_0\omega$, and $Q\xi_0\omega$, respectively.
We expect that this ambiguity does not affect the on-shell physical amplitudes.

\subsection{Four-point amplitudes with external fermions}\label{four point amplitudes}

As evidence that the above Feynman rules work well,
we show that they actually reproduce the on-shell physical 
four-point amplitudes with external fermions. 

The external states are the on-shell states $|V\rangle$ 
and $|\Psi\rangle$ satisfying linearized
equations of motion (\ref{linearized eom}) and
gauge conditions (\ref{gauge conditions}).
Since they have to also satisfy the linearized constraint, 
we can replace $\omega$ and $\xi_0\omega$ connected to external legs 
with the on-shell states $\eta|\Psi\rangle$ and $|\Psi\rangle$, respectively.
The vertices for computing four-point amplitudes can be read from the action
as follows.

The three-boson vertex is obtained from the NS action (\ref{NS action}) as
\begin{equation}
 S_{BBB} = \frac{\kappa}{3!}\langle\eta V,[V,QV]\rangle.\label{3 bosons}
\end{equation}

The boson-fermion-fermion (Yukawa) vertex in (\ref{R action 2}) is given by
\begin{align}
S_{BFF} =&\ \frac{\kappa}{2}\langle\eta\Psi,[V,Q\tilde{\Xi}]\rangle\nonumber\\
\sim&\ \frac{\kappa}{2}\langle\omega,[V,\omega]\rangle,\label{Yukawa}
\end{align}
where the above replacement was made in the second line.

The boson-boson-fermion-fermion coupling
and four-fermion coupling are similarly read from (\ref{R action 2}) and (\ref{R action 4}),
respectively, as
\begin{align}
 S_{BBFF} =&\ \frac{\kappa^2}{4}\left(\langle\eta\Psi,[V,QV,Q\tilde{\Xi}]\rangle
+\langle\eta\Psi,[V,[V,Q\tilde{\Xi}]]\right)\nonumber\\
\sim&\ \frac{\kappa^2}{4}\langle\omega,[V,QV,\omega]\rangle,\label{BBFF}\\ 
S_{FFFF} =&\ \frac{\kappa^2}{4!}\langle\eta\Psi,[\Psi,(Q\tilde{\Xi})^2]\rangle\nonumber\\
\sim&\ \frac{\kappa^2}{4!}\langle\omega,[\xi_0\omega,\omega^2]\rangle
=\frac{\kappa^2}{4!}\langle\xi_0\omega,[\omega^3]\rangle.\label{four fermi}
\end{align}
We used in (\ref{BBFF}) the fact that 
\begin{equation}
 \langle\omega,[V,[V,\omega]]\rangle = \langle[V,\omega],[V,\omega]\rangle 
= 0.
\end{equation}

\begin{figure}[htbp]
\begin{minipage}{0.33\hsize}
 \begin{center}
 \includegraphics[width=5cm]{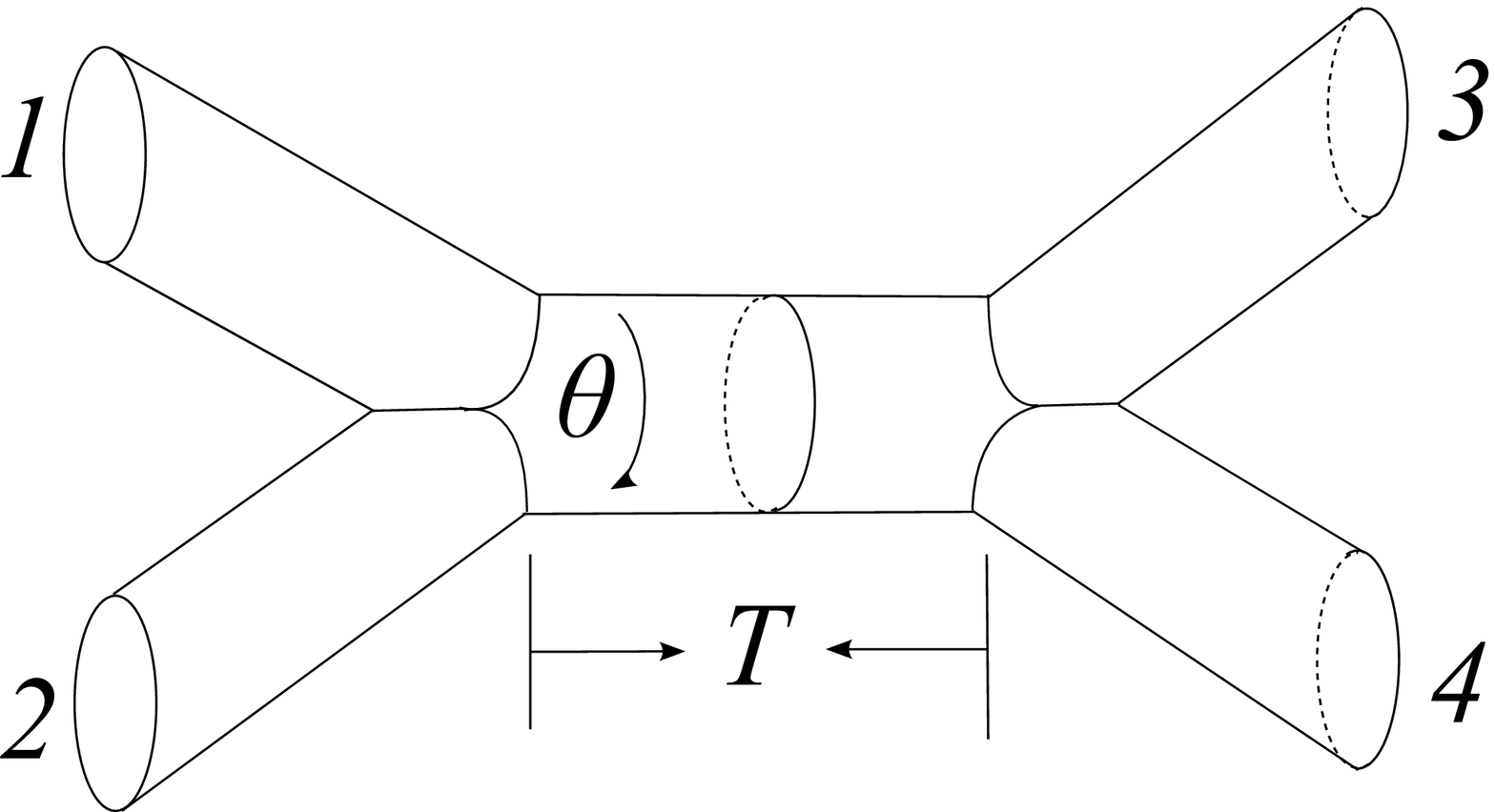}\\
(a)
 \end{center}
\end{minipage}
\begin{minipage}{0.33\hsize}
 \begin{center}
 \includegraphics[width=3.3cm]{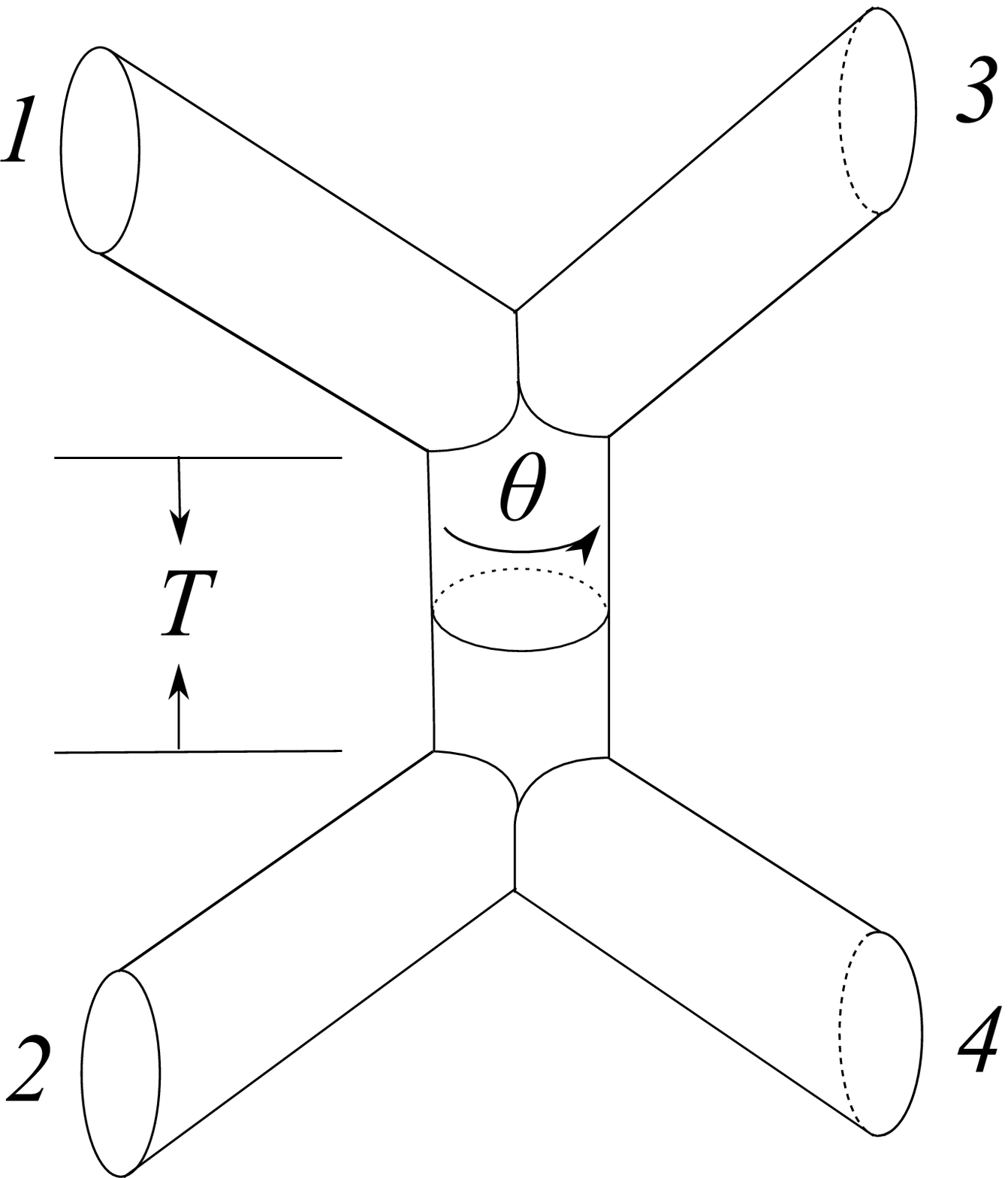}\\
(b)
 \end{center}
\end{minipage}
\begin{minipage}{0.33\hsize}
 \begin{center}
 \includegraphics[width=4.3cm]{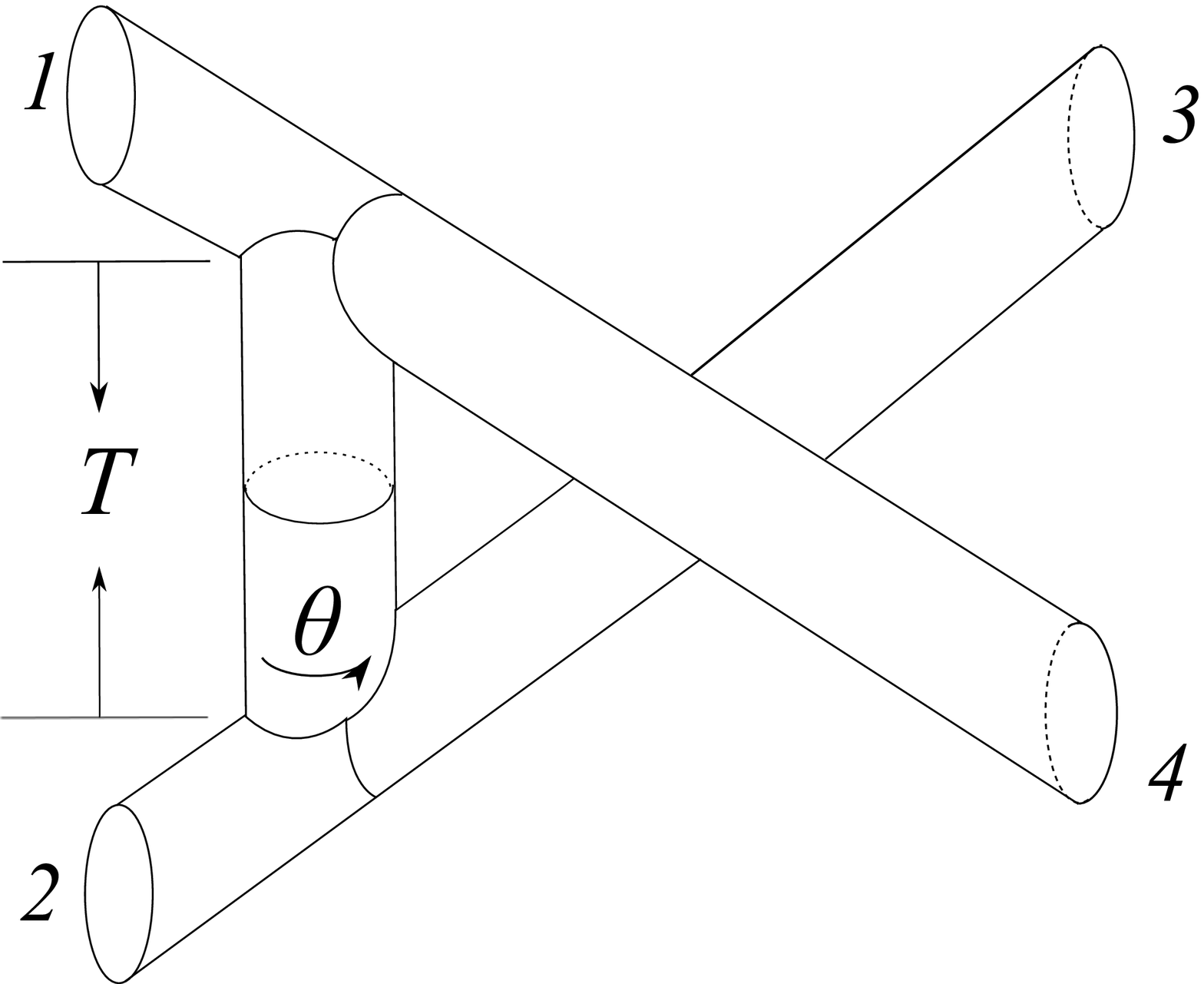} \\
(c)
 \end{center}
\end{minipage}
\vspace{2mm}
\caption{Three Feynman diagrams composed of two three-point vertices and 
a propagator called
(a) $s$-channel, (b) $t$-channel, and (c) $u$-channel diagrams.
The external legs should be read as semi-infinite cylinders. 
The line winding around the propagator denotes the contour
along which $\xi$ and $b^\pm$ are integrated.
}\label{Feynman}
 \end{figure}

\begin{figure}[htbp]
\begin{center}
\includegraphics[width=4cm]{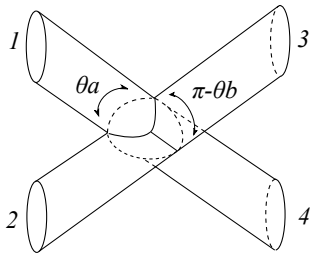} 
\end{center}
\caption{Feynman diagram using the four-point interaction}
\label{Feynman 4 vertex}
\end{figure}

Now let us first calculate the four-fermion amplitude $A_{FFFF}$.
In general, a four-point amplitudes in the heterotic string field theory
has contributions from the four Feynman diagrams depicted in Fig.~\ref{Feynman} 
and Fig.~\ref{Feynman 4 vertex}.
The first three diagrams, in Fig.~\ref{Feynman}, 
the $s$-channel, $t$-channel, and $u$-channel diagrams
for $A_{FFFF}$, can be drawn using the two Yukawa vertices (\ref{Yukawa}) and 
the NS propagator (\ref{NS propagator}). Their contributions to the amplitudes 
are given by
\begin{align}
 A^{stu}_{FFFF} =& \kappa^2 \int_0^\infty \!\!dT\!\!\int_0^{2\pi}\frac{d\theta}{2\pi}
\Bigg(\langle
 (\eta\Psi_1)(\eta\Psi_2)\xi_sb_s^+b_s^-(\eta\Psi_3)(\eta\Psi_4)\rangle_s\nonumber\\
&
+\langle(\eta\Psi_1)(\eta\Psi_3)\xi_tb_t^+b_t^-(\eta\Psi_2)(\eta\Psi_4)\rangle_t
+\langle(\eta\Psi_1)(\eta\Psi_4)\xi_ub_u^+b_u^-(\eta\Psi_2)(\eta\Psi_3)\rangle_u\Bigg),
\label{stu contribution}
\end{align}
where the correlation function $\langle\cdots\rangle_{i=s,t,u}$ is 
evaluated as 
the conformal field theory (CFT) in the large Hilbert space on the 
$i$-channel Feynman diagram. 
The operators $\xi_i$ and $b^\pm_i$ inserted in each term
are integrated along the contour winding around the propagator 
as depicted in Fig.~\ref{Feynman}.
We can remove this $\xi$-insertion using an $\eta$ on an external state, 
for example $\eta\Psi_1$. As a result, the contribution (\ref{stu contribution})  
becomes
\begin{align}
 A^{stu}_{FFFF} =& \kappa^2 \int_0^\infty \!\!dT\!\!\int_0^{2\pi}\frac{d\theta}{2\pi}
\Bigg(\langle
 \Psi_1(\eta\Psi_2)b_s^+b_s^-(\eta\Psi_3)(\eta\Psi_4)\rangle_s\nonumber\\
&\hspace{1cm}
+\langle\Psi_1(\eta\Psi_3)b_t^+b_t^-(\eta\Psi_2)(\eta\Psi_4)\rangle_t
+\langle\Psi_1(\eta\Psi_4)b_u^+b_u^-(\eta\Psi_2)(\eta\Psi_3)\rangle_u\Bigg),\nonumber\\
=&\kappa^2\int_{F_s\cup F_t\cup F_u} d^2z_1 \langle\xi U_F^{(-1/2)}(z_1)V_F^{(-1/2)}(0)
V_F^{(-1/2)}(1)V_F^{(-1/2)}(\infty)\rangle_C,
\end{align}
\begin{figure}[htbp]
\begin{center}
  \includegraphics[width=7cm]{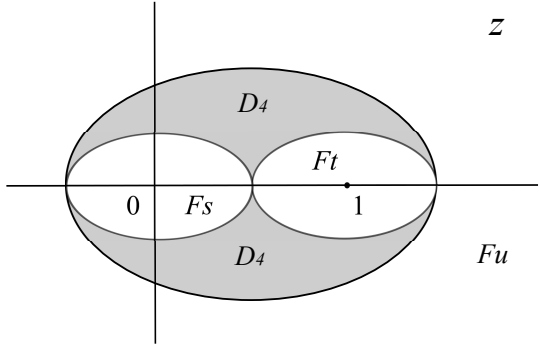}
 \end{center}
\caption{Complex plane}
\label{z plane}
\end{figure}
where, in the second equality, we conformally mapped the Feynman diagram 
to the complex plane\cite{LeClair:1988sp,LeClair:1988sj} depicted 
in Fig.~\ref{z plane}. The external strings, one, two, three and four, are mapped to
$z=z_1$, $0$, $1$, and $\infty$, respectively.
The operator $V_F^{(-1/2)}$ is the conventional on-shell physical 
fermion vertex 
in the $-1/2$-picture\cite{Friedan:1985ge} and related to the on-shell physical R state
$|\Psi\rangle$ as $|\Psi\rangle = \xi_0V_F^{(-1/2)}(0)|0\rangle$.
The integrated vertex operator $U_F$ is also the conventional one: $V_F=c\bar{c}U_F$.
The correlation function $\langle\cdots\rangle_C$ of these operators is evaluated as the CFT 
on the complex plane, where the parameters $\{T,\theta\}$ of the Feynman diagram are mapped into 
the single complex moduli parameter $z_1$. The integration region coming from the $i$-channel
$(i=s,t,u)$ contribution are depicted as $F_i$ in Fig.~\ref{z plane}. 

For $A_{FFFF}$,
the last Feynman diagram depicted in Fig.~\ref{Feynman 4 vertex} 
is drawn using the four-fermi interaction (\ref{four fermi}).
The contribution from this diagram is
\begin{align}
 A_{FFFF}^4 =& \frac{\kappa^2}{4}\int d\theta_a d\theta_b\Bigg(\langle
b_{C_a}b_{C_b}\Psi_1(\eta\Psi_2)(\eta\Psi_3)(\eta\Psi_4)\rangle_4
+\sum_{i=2}^4\langle \Psi(1)\rightarrow \Psi(i)\rangle_4\Bigg)\nonumber\\
=&\kappa^2\int_{D_4} d^2z_1 \langle\xi U_F^{(-1/2)}(z_1)V_F^{(-1/2)}(0)
V_F^{(-1/2)}(1)V_F^{(-1/2)}(\infty)\rangle_C,
\end{align}
where the correlation function in the first line is evaluated as the CFT 
on the diagram in Fig.~\ref{Feynman 4 vertex}. 
The moduli parameters $\{\theta_a,\theta_b\}$ and 
corresponding anti-ghost insertions, $\{b_{C_a},b_{C_b}\}$,
are the same as those given in Ref.~\citen{Kugo:1989tk}. 
In the second equality, we mapped the Feynman diagram to the complex plane and
used the fact that the position of the $\xi(z)$ is irrelevant.\cite{Friedan:1985ge}
This contribution fills the shaded region $D_4$ in Fig.~\ref{z plane}.

By summing up all of these contributions,
the four-fermion amplitude $A_{FFFF}$ is finally given by integrating over 
the complete moduli space $\mathcal{M}_{(0,4)}=F_s\cup F_t\cup F_u\cup D_4$,
the whole complex plane:
\begin{align}
 A_{FFFF} =& A_{FFFF}^{stu} + A_{FFFF}^4\nonumber\\
=& \kappa^2\int_{\mathcal{M}_{(0,4)}} 
d^2z_1 \langle\xi U_F^{(-1/2)}(z_1)V_F^{(-1/2)}(0)V_F^{(-1/2)}(1)
V_F^{(-1/2)}(\infty)\rangle_C.
\label{four fermion}
\end{align}
This coincides with the well-known four-Ramond string 
amplitude.\cite{Friedan:1985ge}

We can similarly show that 
the two-boson-two-fermion amplitude
coincides with the well-known result.
We suppose that the initial strings, strings one and two, are the NS strings 
and the final strings, strings three and four, are the R strings.
In this case, the $s$-channel Feynman diagram consists of the three-boson 
vertex (\ref{3 bosons}),
the Yukawa vertex (\ref{Yukawa}), and the NS propagator (\ref{NS propagator}).
Its contribution to the amplitudes is 
\begin{align}
 A_{BBFF}^s =& \frac{\kappa^2}{2}\int_0^\infty \!\!dT\!\!\int_0^{2\pi}\frac{d\theta}{2\pi}
\Bigg(\langle(\eta V_1)(QV_2)\xi_sb_s^+b_s^-(\eta\Psi_3)(\eta\Psi_4)\rangle_s\nonumber\\
&\hspace{50mm}
+\langle(Q V_1)(\eta V_2)\xi_sb_s^+b_s^-(\eta\Psi_3)(\eta\Psi_4)\rangle_s
\Bigg),\nonumber\\
=& \frac{\kappa^2}{2}
\int_0^\infty \!\!dT\!\!\int_0^{2\pi}\frac{d\theta}{2\pi}\Bigg(
\langle V_1(QV_2)b_s^+b_s^-(\eta\Psi_3)(\eta\Psi_4)\rangle_s\nonumber\\
&\hspace{50mm}
+\langle(Q V_1) V_2b_s^+b_s^-(\eta\Psi_3)(\eta\Psi_4)\rangle_s
\Bigg),\nonumber\\
=& \frac{\kappa^2}{2} \int_{F_s} d^2z_1 
\Big(\langle\xi U_B^{(-1)}(z_1)V_B^{(0)}(0)V_F^{-1/2}(1)V_F^{-1/2}(\infty)\rangle_C\nonumber\\
&\hspace{45mm}
+\langle U_B^{(0)}(z_1) \xi V_B^{(-1)}(0)V_F^{(-1/2)}(1)V_F^{(-1/2)}(\infty)\rangle_C
\Big).\label{boson fermion}
\end{align}
Here, in a similar way to the four-fermion amplitudes, 
we eliminated $\xi$-insertion in the second equality by using 
the $\eta$ on the external NS state $\eta V_1$, 
and mapped the Feynman graph to the complex plane in the third equality. 
The boson vertex operator $V_B^{(-1)}$ is the conventional one 
in the $(-1)$-picture
and related to the on-shell physical NS state $|V\rangle$ as 
$|V\rangle = \xi V_B(0)^{(-1)}|0\rangle$.  
The vertex $V_B^{(0)}$ in the $0$-picture also appears 
in (\ref{boson fermion}) because 
$Q|V\rangle=V_B^{(0)}(0)|0\rangle$.\cite{Friedan:1985ge} 
The integrated vertex $U_B$ is given by $V_B=c\bar{c}U_B$. 

The other two, $t$-channel and $u$-channel, diagrams consist of
the two Yukawa vertices (\ref{Yukawa}) and the R propagator 
(\ref{R propagator}). 
Their contributions are
\begin{align}
 A_{BBFF}^{tu} =& \frac{\kappa^2}{2}
\int_0^\infty \!\!dT\!\!\int_0^{2\pi}\frac{d\theta}{2\pi}
\Bigg(
\langle(\eta V_1)(\eta\Psi_3)\xi_tb_t^+b_t^-(QV_2)(\eta\Psi_4)\rangle_t
+\langle(QV_1)(\eta\Psi_3)\xi_tb_t^+b_t^-(\eta V_2)(\eta\Psi_4)\rangle_t
\nonumber\\
&\hspace{20mm}
+\langle(\eta V_1)(\eta\Psi_4)\xi_ub_u^+b_u^-(QV_2)(\eta\Psi_3)\rangle_u
+\langle(QV_1)(\eta\Psi_4)\xi_ub_u^+b_u^-(\eta V_2)(\eta\Psi_3)\rangle_u
\Bigg),\nonumber\\
\nonumber
\end{align}
\begin{align}
=& \frac{\kappa^2}{2}
\int_0^\infty \!\!dT\!\!\int_0^{2\pi}\frac{d\theta}{2\pi}
\Bigg(
\langle V_1(\eta\Psi_3)b_t^+b_t^-(QV_2)(\eta\Psi_4)\rangle_t
+\langle(QV_1)(\eta\Psi_3)b_t^+b_t^- V_2 (\eta\Psi_4)\rangle_t
\nonumber\\
&\hspace{30mm}
+\langle V_1(\eta\Psi_4)b_u^+b_u^-(QV_2)(\eta\Psi_3)\rangle_u
+\langle(QV_1)(\eta\Psi_4)b_u^+b_u^- V_2 (\eta\Psi_3)\rangle_u
\Bigg),\nonumber\\
=& \frac{\kappa^2}{2}\int_{F_t\cup F_u} d^2z_1 \Bigg(\langle
 \xi U_B^{(-1)}(z_1)V_B^{(0)}(0)V_F^{(-1/2)}(1)V_F^{(-1/2)}(\infty)\rangle_C\nonumber\\
&\hspace{4.5cm}
+\langle U_B^{(0)}(z_1) \xi V_B^{(-1)}(0) V_F^{(-1/2)}(1)V_F^{(-1/2)}(\infty)\rangle_C
\Bigg).
\end{align}

The missing integration region $D_4$ is again filled 
by the contribution of the graph 
consisting of the four-string interaction (\ref{BBFF}),
\begin{align}
 A_{BBFF}^4 =& \frac{\kappa^2}{2}
\int d\theta_a d\theta_b
\Bigg(
\langle b_{C_a}b_{C_b} V_1(z) (QV_2)(\eta\Psi_3)(\eta\Psi_4)\rangle_4
+\langle b_{C_a}b_{C_b}(QV_1) V_2 (\eta\Psi_3)(\eta\Psi_4)\rangle_4
\Bigg),\nonumber\\
=& \frac{\kappa^2}{2} \int_{D_4} d^2z_1 \Bigg(\langle
 \xi U_B^{(-1)}(z_1)V_B^{(0)}(0)V_F^{(-1/2)}(1)V_F^{(-1/2)}(\infty)\rangle_C\nonumber\\
&\hspace{4.5cm}
+\langle U_B^{(0)}(z_1) \xi V_B^{(-1)}(0)V_F^{(-1/2)}(1)V_F^{(-1/2)}(\infty)\rangle_C
\Bigg).
\end{align}

As a result, the two-boson-two-fermion amplitude becomes 
the well-known form as
\begin{align}
 A_{BBFF} =& A_{BBFF}^s + A_{BBFF}^{tu} 
+  A_{BBFF}^4\nonumber\\
\nonumber\\
=& \frac{\kappa^2}{2} \int_{\mathcal{M}_{(0,4)}} d^2z_1 
\Bigg(\langle \xi U_B^{(-1)}(z_1)V_B^{(0)}(0)V_F^{(-1/2)}(1)V_F^{(-1/2)}(\infty)\rangle_C\nonumber\\
&\hspace{3cm}
+\langle U_B^{(0)}(z_1)\xi V_B^{(-1)}(0)V_F^{(-1/2)}(1)V_F^{(-1/2)}(\infty)\rangle_C
\Bigg),\nonumber\\
=& \kappa^2 \int_{\mathcal{M}_{(0,4)}} d^2z_1 
\langle \xi U_B^{(-1)}(z_1)V_B^{(0)}(0)V_F^{(-1/2)}(1)V_F^{(-1/2)}(\infty)\rangle_C,
\end{align}
where in the last equality we used a $\xi$ manipulation,\cite{Friedan:1985ge}
which is only available if the moduli $z_1$ is integrated over the whole complex plane,
$\mathcal{M}_{(0,4)}$.

\section{Open questions and discussion}\label{Discussion}

An important remaining problem is to prove the existence of a non-trivial
solution to Eq.~(\ref{consistency id2}). If it has no non-trivial 
solution at some order in $\Psi$, we must relax the G-ansatz, for example
by allowing the string field $\bar{B}_\eta(V)$ in (\ref{explicit Bs}) 
to appear in the correction terms. 
Then the general form (\ref{EOM}) has to be modified.

A more important task is to give the complete equations of motion and 
the gauge transformations in a closed form.
The general form (\ref{EOM}) determined so as to be invariant 
under the $\Lambda_0$-gauge transformation (\ref{Lambda zero}) 
and consistent with the identity (\ref{id1}), 
or its alternative in the case mentioned above,
will give an important clue in this study.
In particular, the $\Omega$ seems to play an important role 
since it also appears in the constraint
(\ref{constraint}) and can be considered as a kind of non-linear 
extension of the R string field $\eta\Psi$ in the open superstring 
field theory. 

It is also important to find the action giving the complete
equations of motion. In this regard, it is worth noting that,
in contrast to the equations of motion (\textit{i.e.} $\Omega$
and $\Sigma$) (\ref{L omega and sigma}),
(\ref{NL omega and sigma}), and (\ref{NNL omega and sigma}), 
not all the possible terms allowed by the ghost number and
the picture number counting appear in
the action (\ref{R action 4}) and (\ref{R action 6}). 
We conjecture, from the explicit form of $S_R^{(4)}$, (\ref{R action 4}),
and $S_R^{(6)}$, (\ref{R action 6}), that 
each term in the R action includes the same number of 
$\Psi$s and $\Xi$s. Then, the possible terms in $S_R^{(2n+2)}$
take the form,
\begin{equation}
 \kappa^{2n}\langle\eta\Psi,\mspd{\Psi,\qpsi^{k_1},\qxi^{l_1},\mspd{
\Psi,\qpsi^{k_2},\qxi^{l_2},\mspd{\cdots,\mspd{
\Psi,\qpsi^{k_m},\qxi^{l_m}}\cdots}}}\rangle,\nonumber
\end{equation}
or that obtained by exchanging $\eta\Psi$ and one of the $Q_G\Xi$
in the middle ($2\ge i\ge m-1$) for $m\ge 3$,
where $1\le m\le n$ and the non-negative integers $k_i$ and $l_i$
$(1\le i\le m)$ are restricted by $\sum_{i=1}^mk_i=n-m$, 
$\sum_{i=1}^ml_i=n+1$ and $k_m+l_m\ne0$.
Since these forms are much simpler than those of possible terms 
in the equations of motion, given in \S\ref{action with a constraint}, 
it is interesting to look for some guiding principle to determine
the action with the constraint without the help of the equations of motion.

Another remaining problem is to confirm that 
the Feynman rules constructed in \S\S \ref{Feynman rule} reproduce
general tree-level physical amplitudes since the proposed
prescription has an ambiguity as already mentioned.
The possible difficulty due to this ambiguity occurs 
when the $\omega$ in question is connected to the other vertex as
an internal line, which first happens in the five-point amplitudes
with the four external fermions. 
It must be confirmed whether this five-point amplitude is independent of 
the ambiguity or not. We might have to modify the Feynman rules,
which seems to be suggested from the computation of the five-point 
amplitudes with external fermions in open superstring field 
theory.\cite{Michishita:RIKEN}
Rules for computing loop amplitudes should also be found, which
forces us to investigate the gauge-fixing 
procedure\cite{Kroyter:2012ni,Berkovits:2012np} in more detail.

\vspace{5mm}
\section*{Acknowledgments}

The author would like to thank one of the referees for his/her
useful comments and bringing Ref. \citen{Michishita:RIKEN} to his attention.

\appendix

\section{The next-to-next-to-leading order corrections
to the gauge transformations}\label{appendix A}

In this Appendix, we give the explicit forms of 
the next-to-next-to-leading order corrections
$\bar{B}_{\delta^{(6)}}$ and $\delta^{(4)}\Psi$ to
the gauge transformations.

The next-to-next-to-leading order corrections to the $\Lambda_1$-gauge transformation are given by
\begin{align}
  \bar{B}_{\delta^{(6)}_{\Lambda_1}}(V)=&
-\frac{\kappa^6}{6!}\mspd{\Psi,\qpsi^2,\epsi^3,\eLambda}
-\frac{\kappa^6}{6!}\mspd{\mspd{\Psi,Q_G\Psi},\Psi,\epsi^3,\eLambda}\nonumber\\
&-\frac{\kappa^6}{80}\kappa^6\mspd{\mspd{\Psi,\eta\Psi},\Psi,Q_G\Psi,\epsi^2,\eLambda}
-\frac{4}{6!}\kappa^6\mspd{\mspd{\Psi,Q_G\Psi,\eta\Psi},\Psi,\epsi^2,\eLambda}\nonumber\\
&+\frac{4}{6!}\kappa^6\mspd{\mspd{\Psi,\epsi^2},\Psi,Q_G\Psi,\eta\Psi,\eLambda}
-\frac{\kappa^6}{5!}\mspd{\mspd{\Psi,Q_G\Psi,\epsi^2},\Psi,\eta\Psi,\eLambda}\nonumber\\
&+\frac{\kappa^6}{6!}\mspd{\mspd{\Psi,\epsi^3},\Psi,Q_G\Psi,\eLambda}
+\frac{2}{6!}\kappa^6\mspd{\mspd{\Psi,Q_G\Psi,\epsi^3},\Psi,\eLambda}\nonumber\\
&+\frac{2}{6!}\kappa^6\mspd{\Psi,\mspd{\Psi,Q_G\Psi,\epsi^3,\eLambda}}
+\frac{\kappa^6}{6!}\mspd{\Psi,Q_G\Psi,\mspd{\Psi,\epsi^3,\eLambda}}\nonumber\\
&+\frac{\kappa^6}{80}\mspd{\Psi,\eta\Psi,\mspd{\Psi,Q_G\Psi,\epsi^2,\eLambda}}
+\frac{4}{6!}\kappa^6\mspd{\Psi,Q_G\Psi,\eta\Psi,\mspd{\Psi,\epsi^2,\eLambda}}\nonumber\\
&+\frac{\kappa^6}{6!}\mspd{\Psi,\epsi^2,\mspd{\Psi,Q_G\Psi,\eta\Psi,\eLambda}}
+\frac{\kappa^6}{5!}\mspd{\Psi,Q_G\Psi,\epsi^2,\mspd{\Psi,\eta\Psi,\eLambda}}\nonumber\\
&-\frac{\kappa^6}{6!}\mspd{\Psi,\epsi^3,\mspd{\Psi,Q_G\Psi,\eLambda}}
-\frac{5}{6!}\kappa^6\mspd{\mspd{\Psi,\mspd{\Psi,\eta\Psi}},\Psi,\epsi^2,\eLambda}\nonumber\\
&+\frac{\kappa^6}{72}\mspd{\mspd{\Psi,\mspd{\Psi,\epsi^2}},\Psi,\eta\Psi,\eLambda}
-\frac{\kappa^6}{48}\mspd{\mspd{\Psi,\eta\Psi,\mspd{\Psi,\eta\Psi}},\Psi,\eta\Psi,\eLambda}\nonumber\\
&-\frac{\kappa^6}{48}\mspd{\mspd{\Psi,\eta\Psi},\mspd{\Psi,\eta\Psi},\Psi,\eta\Psi,\eLambda}
-\frac{\kappa^6}{6!}\mspd{\mspd{\Psi,\mspd{\Psi,\epsi^3}},\Psi,\eLambda}\nonumber\\
&-\frac{4}{6!}\kappa^6\mspd{\mspd{\Psi,\eta\Psi,\mspd{\Psi,\epsi^2}},\Psi,\eLambda}
+\frac{\kappa^6}{80}\mspd{\mspd{\Psi,\epsi^2,\mspd{\Psi,\eta\Psi}},\Psi,\eLambda}\nonumber\\
&+\frac{5}{6!}\kappa^6\mspd{\mspd{\Psi,\eta\Psi},\mspd{\Psi,\epsi^2},\Psi,\eLambda}
+\frac{5}{6!}\kappa^6\mspd{\Psi,\mspd{\Psi,\eta\Psi},\mspd{\Psi,\epsi^2,\eLambda}}\nonumber\\
&+\frac{\kappa^6}{80}\mspd{\Psi,\mspd{\mspd{\Psi,\eta\Psi},\Psi,\epsi^2,\eLambda}}
+\frac{\kappa^6}{48}\mspd{\Psi,\eta\Psi,\mspd{\Psi,\eta\Psi},\mspd{\Psi,\eta\Psi,\eLambda}}\nonumber\\
&-\frac{4}{6!}\kappa^6\mspd{\Psi,\mspd{\mspd{\Psi,\epsi^2},\Psi,\eta\Psi,\eLambda}}
+\frac{\kappa^6}{24}\mspd{\Psi,\eta\Psi,\mspd{\mspd{\Psi,\eta\Psi},\Psi,\eta\Psi,\eLambda}}\nonumber\\
&-\frac{\kappa^6}{72}\mspd{\Psi,\mspd{\Psi,\epsi^2},\mspd{\Psi,\eta\Psi,\eLambda}}
-\frac{\kappa^6}{6!}\mspd{\Psi,\mspd{\mspd{\Psi,\epsi^3},\Psi,\eLambda}}\nonumber\\
&-\frac{5}{6!}\kappa^6\mspd{\Psi,\eta\Psi,\mspd{\mspd{\Psi,\epsi^2},\Psi,\eLambda}}
+\frac{5}{6!}\kappa^6\mspd{\Psi,\epsi^2,\mspd{\mspd{\Psi,\eta\Psi},\Psi,\eLambda}}\nonumber\\
&-\frac{\kappa^6}{6!}\mspd{\Psi,\mspd{\Psi,\mspd{\Psi,\epsi^3,\eLambda}}}
-\frac{4}{6!}\kappa^6\mspd{\Psi,\mspd{\Psi,\eta\Psi,\mspd{\Psi,\epsi^2,\eLambda}}}\nonumber\\
&-\frac{5}{6!}\kappa^6\mspd{\Psi,\eta\Psi,\mspd{\Psi,\mspd{\Psi,\epsi^2,\eLambda}}}
-\frac{\kappa^6}{5!}\mspd{\Psi,\mspd{\Psi,\epsi^2,\mspd{\Psi,\eta\Psi,\eLambda}}}\nonumber\\
&-\frac{\kappa^6}{48}\mspd{\Psi,\eta\Psi,\mspd{\Psi,\eta\Psi,\mspd{\Psi,\eta\Psi,\eLambda}}}
a+\frac{5}{6!}\kappa^6\mspd{\Psi,\epsi^2,\mspd{\Psi,\mspd{\Psi,\eta\Psi,\eLambda}}},\\
 \delta^{(4)}_{\Lambda_1}\Psi=&
-\frac{\kappa^5}{5!}\mspd{\Psi,\qpsi^2,\epsi^2,\eLambda}
-\frac{\kappa^5}{5!}\mspd{\mspd{\Psi,Q_G\Psi},\Psi,\epsi^2,\eLambda}\nonumber\\
&-\frac{7}{5!}\kappa^5\mspd{\mspd{\Psi,\eta\Psi},\Psi,Q_G\Psi,\eta\Psi,\eLambda}
-\frac{3}{5!}\kappa^5\mspd{\mspd{\Psi,Q_G\Psi,\eta\Psi},\Psi,\eta\Psi,\eLambda}\nonumber\\
&+\frac{\kappa^5}{5!}\mspd{\mspd{\Psi,\epsi^2},\Psi,Q_G\Psi,\eLambda}
-\frac{3}{5!}\kappa^5\mspd{\mspd{\Psi,Q_G\Psi,\epsi^2},\Psi,\eLambda}\nonumber\\
&+\frac{2}{5!}\kappa^5\mspd{\Psi,\mspd{\Psi,Q_G\Psi,\epsi^2,\eLambda}}
+\frac{\kappa^5}{5!}\mspd{\Psi,Q_G\Psi,\mspd{\Psi,\epsi^2,\eLambda}}\nonumber\\
&+\frac{\kappa^5}{3\cdot 5!}\mspd{\Psi,\eta\Psi,\mspd{\Psi,Q_G\Psi,\eta\Psi,\eLambda}}
+\frac{3}{5!}\kappa^5\mspd{\Psi,Q_G\Psi,\eta\Psi,\mspd{\Psi,\eta\Psi,\eLambda}}\nonumber\\
&-\frac{\kappa^5}{5!}\mspd{\Psi,\epsi^2,\mspd{\Psi,Q_G\Psi,\eLambda}}
-\frac{4}{5!}\kappa^5\mspd{\mspd{\Psi,\mspd{\Psi,\eta\Psi}},\Psi,\eta\Psi,\eLambda}\nonumber\\
&+\frac{4}{5!}\kappa^5\mspd{\mspd{\Psi,\mspd{\Psi,\epsi^2}},\Psi,\eLambda}
-\frac{\kappa^5}{15}\mspd{\mspd{\Psi,\eta\Psi,\mspd{\Psi,\eta\Psi}},\Psi,\eLambda}\nonumber\\
&-\frac{\kappa^5}{15}\mspd{\mspd{\Psi,\eta\Psi},\mspd{\Psi,\eta\Psi},\Psi,\eLambda}
+\frac{7}{5!}\kappa^5\mspd{\Psi,\mspd{\mspd{\Psi,\eta\Psi},\Psi,\eta\Psi,\eLambda}}\nonumber\\
&+\frac{4}{5!}\kappa^5\mspd{\Psi,\mspd{\Psi,\eta\Psi},\mspd{\Psi,\eta\Psi,\eLambda}}
-\frac{\kappa^5}{5!}\mspd{\Psi,\mspd{\mspd{\Psi,\epsi^2},\Psi,\eLambda}}\nonumber\\
&+\frac{\kappa^5}{45}\mspd{\Psi,\eta\Psi,\mspd{\mspd{\Psi,\eta\Psi},\Psi,\eLambda}}
-\frac{\kappa^5}{5!}\mspd{\Psi,\mspd{\Psi,\mspd{\Psi,\epsi^2,\eLambda}}}\nonumber\\
&-\frac{3}{5!}\kappa^5\mspd{\Psi,\mspd{\Psi,\eta\Psi,\mspd{\Psi,\eta\Psi,\eLambda}}}
r+\frac{\kappa^5}{45}\mspd{\Psi,\eta\Psi,\mspd{\Psi,\mspd{\Psi,\eta\Psi,\eLambda}}},
\end{align}

The same order corrections to the $\Lambda_{\frac{1}{2}}$-gauge
transformations are obtained as
\begin{align}
  \bar{B}_{\delta^{(6)}_{\Lambda_{1/2}}}(V)=&
\frac{2}{3\cdot 5!}\kappa^5\mspd{\Psi,Q_G\Psi,\epsi^3,\qLambda}
-\frac{\kappa^5}{5!}\mspd{\mspd{\Psi,Q_G\Psi,\epsi^3},\Lambda_{\frac{1}{2}}}\nonumber\\
&+\frac{\kappa^5}{5!}\mspd{\mspd{\Psi,\mspd{\Psi,\epsi^3}},\Lambda_{\frac{1}{2}}}
+\frac{4}{5!}\kappa^5\mspd{\mspd{\Psi,\eta\Psi,\mspd{\Psi,\epsi^2}},\Lambda_{\frac{1}{2}}}\nonumber\\
&-\frac{4}{5!}\kappa^5\mspd{\mspd{\Psi,\epsi^2,\mspd{\Psi,\eta\Psi}},\Lambda_{\frac{1}{2}}}
+\frac{\kappa^5}{48}\mspd{\mspd{\Psi,\eta\Psi},\Psi,\epsi^2,\qLambda}\nonumber\\
&-\frac{\kappa^5}{36}\mspd{\mspd{\Psi,\epsi^2},\Psi,\eta\Psi,\qLambda}
-\frac{\kappa^5}{144}\mspd{\mspd{\Psi,\epsi^3},\Psi,\qLambda}\nonumber\\
&-\frac{2}{3\cdot 5!}\kappa^5\mspd{\Psi,\mspd{\Psi,\epsi^3,\qLambda}}
-\frac{\kappa^5}{48}\mspd{\Psi,\eta\Psi,\mspd{\Psi,\epsi^2,\qLambda}}\nonumber\\
&+\frac{\kappa^5}{72}\mspd{\Psi,\epsi^2,\mspd{\Psi,\eta\Psi,\qLambda}}
+\frac{\kappa^5}{144}\mspd{\Psi,\epsi^3,\mspd{\Psi,\qLambda}},\\
\delta^{(4)}_{\Lambda_{1/2}}\Psi=&
\frac{\kappa^4}{40}\mspd{\Psi,Q_G\Psi,\epsi^2,\qLambda}\nonumber\\
&+\frac{\kappa^4}{15}\mspd{\mspd{\Psi,\eta\Psi},\Psi,\eta\Psi,\qLambda}
-\frac{\kappa^4}{30}\mspd{\mspd{\Psi,\epsi^2},\Psi,\qLambda}\nonumber\\
&-\frac{\kappa^4}{40}\mspd{\Psi,\mspd{\Psi,\epsi^2,\qLambda}}
+\frac{2}{45}\kappa^4\mspd{\Psi,\eta\Psi,\mspd{\Psi,\eta\Psi,\qLambda}}\nonumber\\
& +\frac{\kappa^4}{30}\mspd{\Psi,\epsi^2,\mspd{\Psi,\qLambda}},
\end{align}

Those to the $\Lambda_{3/2}$-gauge transformations are
\begin{align}
  \bar{B}_{\delta^{(6)}_{\Lambda_{3/2}}}(V) =&
-\frac{2}{6!}\kappa^5\mspd{\Psi,Q_G\Psi,\epsi^3,\eLambdaf}
+\frac{2}{6!}\kappa^5\mspd{\Psi,\mspd{\Psi,\epsi^3,\eLambdaf}}\nonumber\\
&+\frac{\kappa^5}{80}\mspd{\Psi,\eta\Psi,\mspd{\Psi,\epsi^2,\eLambdaf}}
+\frac{\kappa^5}{6!}\mspd{\Psi,\epsi^2,\mspd{\Psi,\eta\Psi,\eLambdaf}}\nonumber\\
&-\frac{\kappa^5}{6!}\mspd{\Psi,\epsi^3,\mspd{\Psi,\eLambdaf}}
-\frac{\kappa^5}{80}\mspd{\mspd{\Psi,\eta\Psi},\Psi,\epsi^2,\eLambdaf}\nonumber\\
&+\frac{4}{6!}\kappa^5\mspd{\mspd{\Psi,\epsi^2},\Psi,\eta\Psi,\eLambdaf}
+\frac{\kappa^5}{6!}\mspd{\mspd{\Psi,\epsi^3},\Psi,\eLambdaf},\\
  \delta^{(4)}_{\Lambda_{3/2}}\Psi =& 
-\frac{2}{5!}\kappa^4\mspd{\Psi,Q_G\Psi,\epsi^2,\eLambdaf}
+\frac{2}{5!}\kappa^4\mspd{\Psi,\mspd{\Psi,\epsi^2,\eLambdaf}}\nonumber\\
&+\frac{\kappa^4}{3\cdot 5!}\mspd{\Psi,\eta\Psi,\mspd{\Psi,\eta\Psi,\eLambdaf}}
-\frac{\kappa^4}{5!}\mspd{\Psi,\epsi^2,\mspd{\Psi,\eLambdaf}}\nonumber\\
&-\frac{7}{5!}\kappa^4\mspd{\mspd{\Psi,\eta\Psi},\Psi,\eta\Psi,\eLambdaf}
+\frac{\kappa^4}{5!}\mspd{\mspd{\Psi,\epsi^2},\Psi,\eLambdaf}.
\end{align}

\vspace{5mm}
\section{Derivation of Eq.~(\ref{EOM})}\label{appendix B}

Let us start to consider the correction to the equations of motion
\begin{align}
 \eta G &=0,\\
 Q_G\eta\Psi &=0.
\end{align}
Since the left-hand sides of these equations are identically annihilated by $Q_G$,
the correction terms have to be also annihilated by $Q_G$ at least up to the equations
of motion. We can classify the correction terms into two categories depending on
whether they are annihilated by $Q_G$ without using the equations of motion or not.
Since the terms, composed of general off-shell string fields, identically annihilated 
by $Q_G$ have to be $Q_G$-exact, let us write them $Q_G\Sigma$ and 
$Q_G(\Omega-\eta\Psi)$ for the NS and the R equations of motion, respectively. 
Then the full equations of motion become the form
\begin{subequations} \label{app EOM}
\begin{align}
 \eta G+Q_G\Sigma + E_{NS} &= 0,\label{app eom NS}\\
 Q_G\Omega + E_R &= 0,\label{app eom R}
\end{align} 
\end{subequations}
where $E_{NS}$ and $E_R$ are the correction terms in the second category,
which have to be self-consistently determined in such a way that $Q_GE_{NS}$ 
and $Q_GE_R$ are proportional to (the left-hand side of) these equations of 
motion (\ref{app EOM}) themselves.

Now let us first determine $E_R$. Under the G-ansatz, we can suppose,
without loss of generality, that 
the possible terms in $E_R$ take the form of 
those obtained by replacing one of the $\Psi$s in the general possible terms 
in $\Omega$, given in \S\ref{eom including R sector}, with $Q_G\Psi$. 
One can deduce that $Q_GE_R$ must be proportional to 
the R equation of motion (\ref{app eom R}) from the form of these general terms.
Then, from the (non-) linearity of $Q_G$ acting on the two (more than two) string products,
the possible form of $E_R$ has to be
\begin{equation}
E_R = \mspd{\Omega, Q_G A} + \mspd{\mspd{\Omega, A}, Q_G A}+\cdots,
\end{equation}
where $A$ is a string field with Grassmann odd and the ghost and picture numbers 
$(G,P)=(1,0)$. The terms represented by dots can be iteratively determined so that
$Q_GE_R$ is proportional to the R equation of motion (\ref{app eom R}). 
We cannot, however, construct such a string field $A$ 
with the desired ghost and picture numbers, and hence $E_R\equiv0$.
We can similarly determine the $E_{NS}$, using the fact that $E_R\equiv0$, as　
\begin{equation}
 E_{NS} = -\frac{\kappa}{2}\mspd{\Omega^2},
\end{equation}
where the numerical coefficient is fixed to reproduce the leading correction term
(\ref{leading eom NS}). 

In conclusion, consistency with the identity (\ref{id1}) requires the full equations
of motion to be in the form of (\ref{EOM}) under the G-ansatz.

\vspace{10mm}
\section{Derivation of Eq.~(\ref{sp expansion}) and the explicit forms of
$\bar{B}^{[2]}_{\delta}$}\label{appendix C}

From the forms of the fundamental possible terms (\ref{possible term}),
the terms built with the one string product in the $\Omega^{(2n+1)}$
and the $\Sigma^{(2n+2)}$
and the terms with two string products in the $\Sigma^{(2n+2)}$ 
can be written for $n\ge1$ as 
\begin{align}
\Omega^{(2n+1)}=& \kappa^{2n}f_n\mspd{\Psi,\qpsi^{n-1},\epsi^{n+1}}+\cdots,\\
 \Sigma^{(2n+2)}=& \kappa^{2n+1}g_n\mspd{\Psi,\qpsi^{n-1},\epsi^{n+2}}\nonumber\\
&+\kappa^{2n+1}\underset{(m,l)\ne(0,0)}{\sum_{m=0}^{n-2}\sum_{l=0}^{n+2}}
g_{n,m,l}\mspd{\Psi,\qpsi^{n-m-2},\epsi^{n-l+2},\mspd{\Psi,\qpsi^m,\epsi^l}}
+\cdots,
\end{align}
where $f_n$, $g_n$ and $g_{n,m,l}$ are the numerical 
coefficients to be determined and 
the dots represent the terms with greater numbers of string products.
The terms built with the two string products in the $\Sigma^{(2n+2)}$ 
only exist for $n\ge2$.
The consistency equation (\ref{consistency id2}) requires, 
neglecting the terms with more than two string products,
that the following equations hold:
\begin{align}
&f_1=-4g_1,\qquad g_1=\frac{1}{4!},\nonumber\\
&\comb{n-1}{m}\comb{n+3}{l}g_n+g_{n,m,l-1}-g_{n,m-1,l}\nonumber\\
&\hspace{5cm}
+2(n-m+1)g_{n-m}f_m\delta_{l,m+1}
-g_{n-m-1}g_m\delta_{l,m+2}=0,\nonumber\\
&\hspace{7.5cm}
(n\ge2,\ 0\le m\le n-2,\ 1\le l\le n+2),\nonumber\\
&\comb{n+3}{l}g_n-g_{n,n-2,l}+\frac{1}{3!}f_{n-1}\delta_{l,n}-\frac{1}{2}g_{n-1}\delta_{l,n+1}
+f_n\delta_{l,n+2}=0,\nonumber\\
&\hspace{10cm}
(n\ge3,\ 1\le l\le n+2),\nonumber\\
&\comb{n-1}{m+1}g_n-g_{n,m,0}=0,\hspace{3cm} (n\ge3,\ 1\le m\le n-2),\nonumber\\
&\comb{n-1}{m}g_n+g_{n,m,n+2}=0,\hspace{2cm} (n\ge3,\ 1\le m\le n-2).
\end{align}
All the coefficients $f_n$, $g_n$ and $g_{n,m,l}$ 
are uniquely determined by solving these equations as
\begin{align}
 &f_n=-\frac{1}{(2n+1)!},\qquad
 g_n=\frac{1}{(2n+2)!},\nonumber\\
 &g_{n,m,l}=
\begin{cases}
\displaystyle
\sum_{k=0}^l\comb{n-1}{m+l+1-k}\comb{n+3}{k}\frac{1}{(2n+2)!}, & \textrm{for}\ 0\le l\le m+1, \\
\displaystyle
-\sum_{k=0}^m\comb{n-1}{k}\comb{n+3}{m+l+1-k}\frac{1}{(2n+2)!}, & \textrm{for}\  m+2\le l\le n+2,
\end{cases}\label{g nml}
\end{align}
with the help of the formula:
\begin{equation}
 \sum_{r=0}^p\comb{n}{r}\comb{m}{p-r}=\comb{n+m}{p}\ \textrm{for}\ p\le n+m.
\end{equation}

The explicit forms of the terms with two string products,
$\bar{B}^{[2]}_{\delta}(V)$, in the gauge transformations of $V$ are given by
\begin{align}
 \bar{B}^{[2]}_{\delta_{\Lambda_1}}(V)=&
-\sum_{n=3}^\infty\sum_{m=1}^{n-2}\sum_{l=0}^{n+1}\left(g_{n,m,l}+\comb{n-1}{m}
\comb{n+2}{l}g_n\right)\nonumber\\
&\hspace{1.5cm}
\times\kappa^{2n+2}\mspd{\Psi,\qpsi^{n-m-1},\epsi^{n-l+1},\mspd{\Psi,\qpsi^m,\epsi^l,\eLambda}}\nonumber\\
&+\sum_{n=2}^\infty\sum_{l=1}^{n+1}\comb{n+2}{l+1}g_n\kappa^{2n+2}
\mspd{\Psi,\qpsi^{n-1},\epsi^{n-l+1},\mspd{\Psi,\epsi^l,\eLambda}}\nonumber\\
&-\sum_{n=3}^\infty\sum_{l=0}^{n+1}\comb{n+2}{l}g_n\kappa^{2n+2}
\mspd{\Psi,\epsi^{n-l+1},\mspd{\Psi,\qpsi^{n-1},\epsi^l,\eLambda}}\nonumber\\
&+\sum_{n=3}^\infty\sum_{m=1}^{n-2}g_{n-m-1}g_m\nonumber\\
&\hspace{1.5cm}
\times\kappa^{2n+2}\mspd{\Psi,\qpsi^{n-m-1},\epsi^{n-m},\mspd{\Psi,\qpsi^m,\epsi^{m+1},\eLambda}}\nonumber\\
&-\sum_{n=3}^\infty\sum_{m=1}^{n-2}(n-m)g_{n-m}f_m\nonumber\\
&\hspace{1.5cm}
\times\kappa^{2n+2}\mspd{\Psi,\qpsi^{n-m-1},\epsi^{n-m+1},\mspd{\Psi,\qpsi^m,\epsi^m,\eLambda}}\nonumber\\
&-\sum_{n=1}^\infty f_n\kappa^{2n+2}\mspd{\Psi,\mspd{\Psi,\qpsi^{n-1},\epsi^{n+1},\eLambda}}\nonumber\\
&+\sum_{n=1}^\infty\frac{1}{2}g_{n-1}\kappa^{2n+2}\mspd{\Psi,\eta\Psi,\mspd{\Psi,\qpsi^{n-1},
\epsi^n,\eLambda}}\nonumber\\
&-\sum_{n=3}^\infty \frac{1}{4!}f_{n-1}\kappa^{2n+2}
\mspd{\Psi,\epsi^2,\mspd{\Psi,\qpsi^{n-1},\epsi^{n-1},\eLambda}}\nonumber\\
&-\frac{3}{4!}\kappa^4\mspd{\Psi,\mspd{\Psi,\epsi^2,\eLambda}}
-\frac{3}{4!}\kappa^4\mspd{\Psi,\eta\Psi,\mspd{\Psi,\eta\Psi,\eLambda}}\nonumber\\
&-\frac{4}{6!}\kappa^6\mspd{\Psi,\mspd{\Psi,Q_G\Psi,\epsi^3,\eLambda}}
-\frac{\kappa^6}{5!}\mspd{\Psi,\eta\Psi,\mspd{\Psi,Q_G\Psi,\epsi^2,\eLambda}}\nonumber\\
&+\frac{\kappa^6}{6!}\mspd{\Psi,\epsi^2,\mspd{\Psi,Q_G\Psi,\eta\Psi,\eLambda}}
-\frac{\kappa^6}{6!}\mspd{\Psi,\epsi^3,\mspd{\Psi,Q_G\Psi,\eLambda}}\nonumber\\
&-\sum_{n=3}^\infty\sum_{m=1}^{n-2}\sum_{l=0}^{n+1}\left(g_{n,m,l}+
\comb{n-1}{m}\comb{n+2}{l}g_n\right)\nonumber\\
&\hspace{1.5cm}
\times\kappa^{2n+2}\mspd{\mspd{\Psi,\qpsi^m,\epsi^l},\Psi,\qpsi^{n-m-1},\epsi^{n-l+2},\eLambda}\nonumber\\
&+\sum_{n=2}^\infty\sum_{l=1}^{n+1}\comb{n+2}{l+1}g_n\kappa^{2n+2}
\mspd{\mspd{\Psi,\epsi^l},\Psi,\qpsi^{n-1},\epsi^{n-l+1},\eLambda}\nonumber\\
&-\sum_{n=3}^\infty\sum_{l=0}^{n+1}g_n\kappa^{2n+2}
\mspd{\mspd{\Psi,\qpsi^{n-1},\epsi^l},\Psi\epsi^{n-l+1},\eLambda}\nonumber\\
&-\sum_{n=2}^\infty\frac{1}{2}g_{n-1}\kappa^{2n+2}
\mspd{\mspd{\Psi,\eta\Psi},\Psi,\qpsi^{n-1},\epsi^n,\eLambda}\nonumber\\
&-\frac{3}{4!}\kappa^4\mspd{\mspd{\Psi,\eta\Psi},\Psi,\eta\Psi,\eLambda}
+\frac{\kappa^4}{4!}\mspd{\mspd{\Psi,\epsi^2},\Psi,\eLambda}\nonumber\\
&-\frac{\kappa^6}{6!}\mspd{\mspd{\Psi,Q_G\Psi},\Psi,\epsi^3,\eLambda}
-\frac{4}{6!}\kappa^6\mspd{\mspd{\Psi,Q_G\Psi,\eta\Psi},\Psi,\epsi^2,\eLambda}\nonumber\\
&-\frac{\kappa^6}{5!}\mspd{\mspd{\Psi,Q_G\Psi,\epsi^2},\Psi,\eta\Psi,\eLambda}
+\frac{2}{6!}\kappa^6\mspd{\mspd{\Psi,Q_G\Psi,\epsi^3},\Psi,\eLambda},
\end{align}
for the $\Lambda_1$-gauge transformation,
\begin{align}
 \bar{B}^{[2]}_{\delta_{\Lambda_{1/2}}}(V)=&
\sum_{n=1}^\infty f_n\kappa^{2n+1}
\mspd{\mspd{\Psi,\qpsi^{n-1},\epsi^{n+1}},\Lambda_{\frac{1}{2}}}\nonumber\\
&+\sum_{n=2}^\infty\sum_{m=0}^{n-2}\sum_{l=0}^{n+1}(n-l+2)g_{n,m,l}\nonumber\\
&\hspace{1.5cm}
\times\kappa^{2n+1}\mspd{\mspd{\Psi,\qpsi^m,\epsi^l},\Psi,\qpsi^{n-m-2},\epsi^{n-l+1},\qLambda}\nonumber\\
&+\sum_{n=2}^\infty\sum_{m=0}^{n-2}\sum_{l=0}^{n+2}lg_{n,m,l}\nonumber\\
&\hspace{1.5cm}
\times\kappa^{2n+1}\mspd{\Psi,\qpsi^{n-m+2},\epsi^{n-l+2},\mspd{\Psi,\qpsi^m,\epsi^{l-1},
\qLambda}}\nonumber\\
&-\sum_{n=2}^\infty\sum_{m=0}^{n-2}(n-m+1)(m+2)g_{n-m-1}f_{m+1}\nonumber\\
&\hspace{1.5cm}
\times\kappa^{2n+1}\mspd{\Psi,\qpsi^{n-m+2},\epsi^{n-m},\mspd{\Psi,\qpsi^m,\epsi^{m+1},\qLambda}},
\end{align}
for the $\Lambda_{1/2}$-gauge transformation, and
\begin{align}
 \bar{B}_{\delta_{\Lambda_{3/2}}}^{[2]}(V)=&
-\sum_{n=3}^\infty\sum_{m=1}^{n-2}\sum_{l=0}^{n+1}\left(
(n-m-1)g_{n,m,l}+(n-1)\comb{n-2}{m}\comb{n+2}{l}g_n\right)\nonumber\\
&\hspace{1.5cm}
\times\kappa^{2n+1}\mspd{\mspd{\Psi,\qpsi^m,\epsi^l},\Psi,\qpsi^{n-m-2},\epsi^{n-l+2},\eLambdaf}\nonumber\\
&+\sum_{n=2}^{\infty}\sum_{l=1}^{n+1}(n-1)\comb{n+2}{l+1}g_n\nonumber\\
&\hspace{1.5cm}
\times\kappa^{2n+1}\mspd{\mspd{\Psi,\epsi^l},\Psi,\qpsi^{n-2},\epsi^{n-l+1},\eLambdaf}\nonumber\\
&-\sum_{n=2}^\infty\frac{1}{2}(n-1)g_{n-1}\kappa^{2n+1}\mspd{\mspd{\Psi,\eta\Psi},\Psi,\qpsi^{n-2},
\epsi^n,\eLambdaf}\nonumber\\
&-\sum_{n=3}^\infty\sum_{m=0}^{n-3}\sum_{l=1}^{n+2}\left((m+1)g_{n,m,l}-(n-1)\comb{n-2}{m}
\comb{n+2}{l}g_n\right)\nonumber\\
&\hspace{1.5cm}
\times\kappa^{2n+1}\mspd{\Psi,\qpsi^{n-m-2},\epsi^{n-l+2},\mspd{\Psi,\qpsi^m,\epsi^{l-1},\eLambdaf}}\nonumber\\
&+\sum_{n=3}^\infty\sum_{m=0}^{n-3}(n-m+1)(m+1)g_{n-m-1}f_{m+1}\nonumber\\
&\hspace{1.5cm}
\times\kappa^{2n+1}\mspd{\Psi,\qpsi^{n-m-2},\epsi^{n-m},\mspd{\Psi,\qpsi^m,\epsi^{m+1},\eLambdaf}}\nonumber\\
&-\sum_{n=2}^\infty\sum_{l=1}^{n+2}(n-1)\comb{n+2}{l-1}g_n\nonumber\\
&\hspace{1.5cm}
\times\kappa^{2n+1}\mspd{\Psi,\epsi^{n-l+2},\mspd{\Psi,\qpsi^{n-2},\epsi^{l-1},\eLambdaf}}\nonumber\\
&-\sum_{n=2}^\infty(n-1)f_n\kappa^{2n+1}\mspd{\Psi,\mspd{\Psi,\qpsi^{n-2},\epsi^{n+1},
\eLambdaf}}\nonumber\\
&+\sum_{n=2}^\infty\frac{1}{2}(n-1)g_{n-1}\kappa^{2n+1}\mspd{\Psi,\eta\Psi,\mspd{\Psi,\qpsi^{n-2},
\epsi^n,\eLambdaf}}\nonumber\\
&-\sum_{n=2}^\infty\frac{1}{4!}(n-1)f_{n-1}\kappa^{2n+1}\mspd{\Psi,\epsi^2,\mspd{\Psi,
\qpsi^{n-2},\epsi^{n-1},\eLambdaf}},
\end{align}
for the $\Lambda_{3/2}$-gauge transformation.

\vspace{10mm}

\end{document}